\documentclass[journal,comsoc,final]{IEEEtran}

\usepackage[noadjust]{cite}
\usepackage{array}
\usepackage{ragged2e}
\usepackage{booktabs}
\usepackage{cellspace, tabularx}
\newcolumntype{P}[1]{>{\centering\arraybackslash}p{#1}}
\newcolumntype{M}[1]{>{\centering\arraybackslash}m{#1}}
\newcolumntype{L}[1]{>{\centering\arraybackslash}l{#1}}
\newcolumntype{Z}{>{\centering\let\newline\\\arraybackslash\hspace{0pt}}X}

\newcolumntype{Y}[1]{>{\raggedright\let\newline\\\arraybackslash\hspace{0pt}}X{#1}}

\usepackage{hhline}
\usepackage{makecell}
\usepackage{multirow}

\usepackage{pifont}

\usepackage{blindtext}
\usepackage{graphicx}
\usepackage{cite}
\usepackage{enumerate}
\usepackage{multirow}
\usepackage{amsmath}
\usepackage[hyphens]{url}
\usepackage[hidelinks]{hyperref}
\hypersetup{breaklinks=true}
\usepackage{dingbat}
\usepackage{xcolor} 
     
\title{A Comprehensive Survey of Interface Protocols for Software Defined Networks}
\author{
	Zohaib Latif,
	Kashif Sharif,
	Fan Li,
	Md Monjurul Karim, and 
	Yu Wang
	
	\thanks{
		Z. Latif, K. Sharif, F. Li, and Md Monjurul Karim are with Beijing Institute of Technology, Beijing, China. (e-mail: z.latif@bit.edu.cn; kashif@bit.edu.cn; fli@bit.edu.cn; mkarim@bit.edu.cn)
	}
	\thanks{
		Y. Wang is with University of North Carolina and Charlotte, NC, USA. (email: yu.wang@uncc.edu)	
	}
}

\begin{document}
\maketitle

\begin{abstract}
Software Defined Networks has seen tremendous growth and deployment in different types of networks. Compared to traditional networks it decouples the control logic from network layer devices, and centralizes it for efficient traffic forwarding and flow management across the domain. This multi-layered architecture has data forwarding devices at the bottom in data plane, which are programmed by controllers in the control plane. The high level application or management plane interacts with control layer to program the whole network and enforce different policies. The interaction among these layers is done through interfaces which work as communication/programming protocols. In this survey, we present a comprehensive study of such interfaces available for southbound, northbound, and east/westbound communication. We have classified each type into different categories based on their properties and capabilities. Virtualization of networks devices is a common practice in Software Defined Networks. This paper also analyzes interface solution which work with different virtualization schemes. In addition, the paper highlights a number of short term and long term research challenges and open issues related to SDN interfaces. 
\end{abstract}


\section{Introduction}
\IEEEPARstart{O}{ver} 4 billion Internet users are connected through almost 500,000 Autonomous Systems (ASes) throughout the world, and still increasing rapidly every year \cite{Users,F2}. Every AS requires a set of applications to manage these networks. Implementation of this diverse range of applications is becoming difficult by using traditional network elements. These elements are usually based on Application Specific Integrated Circuits (ASICs), which may be vendor specific, and requires embedded OS with hundreds of lines of code in low level languages. Configuration and implementation of policies on these devices is not only time consuming but also difficult. Furthermore, it introduces rigidity in networks, due to application specific nature of devices, which reduces network optimization, as well as management. 

Software Defined Networks (SDN) is an emerging form of networks which promises to resolve these issues by decoupling control plane from data plane and provides a software-based centralized controller. By this separation of control plane and data plane, network switches become simple forwarding devices. Whereas, decision making is shifted to controller, which provides a global view of the network and programming abstractions. This centralized entity provides a programmatic control of whole network and provides real-time control of underlying devices to network operators. By using SDN, network management becomes straightforward and helps to remove rigidity from network.

\begin{figure}
	\centering
	\includegraphics[width=0.95\linewidth]{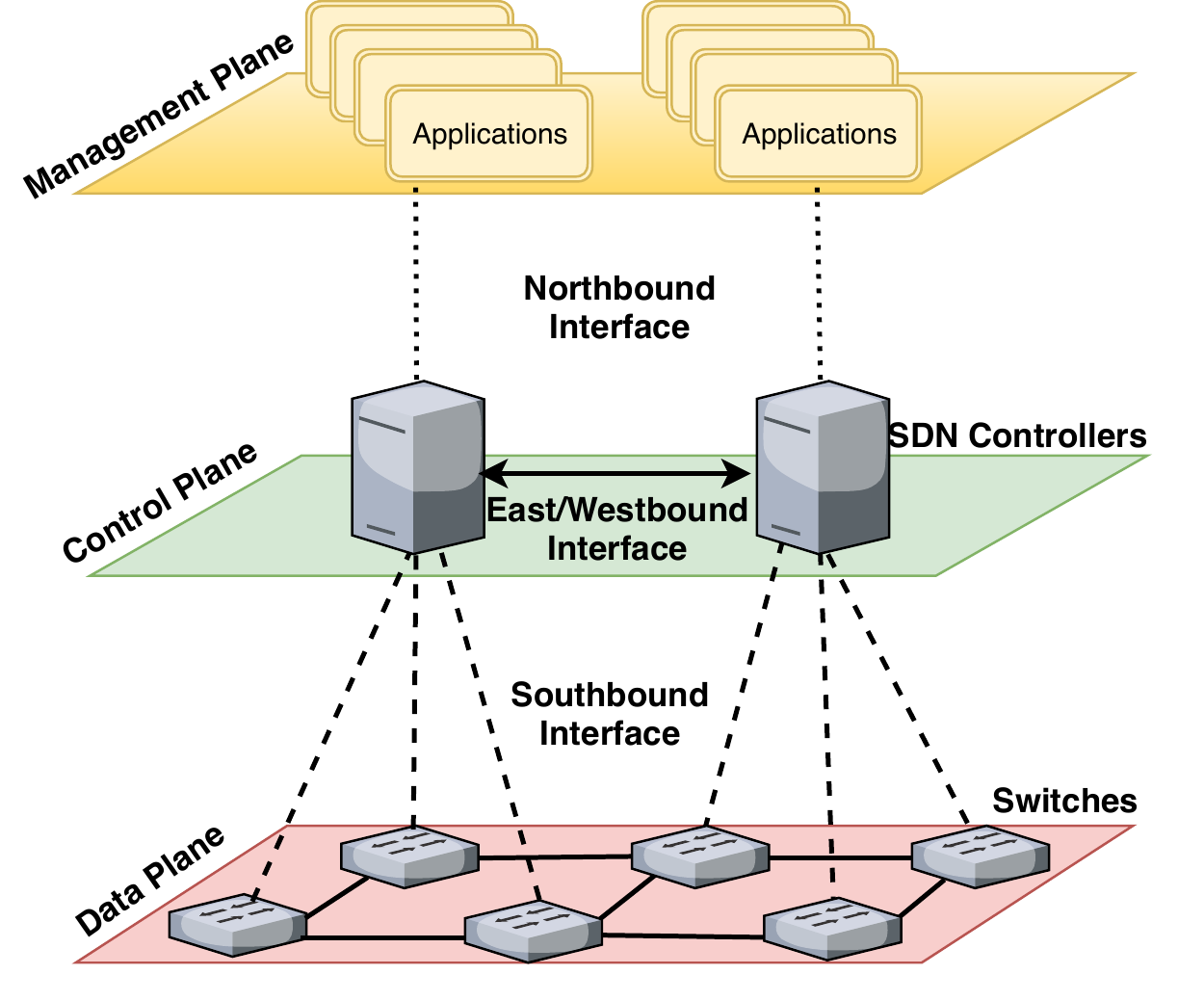}
	\caption{Layered view of SDN Architecture}
	\label{fig:1}
\end{figure}

Fig. \ref{fig:1} represents the layered view of SDN architecture  which has three major planes, referred as; data plane, control plane, and management plane. Data plane contains physical network elements, and these elements form the data path. Control plane has a Network Operating System (NOS), also referred to as controller, which implements rules on data plane devices. These rules and policies are designed in management plane of SDN architecture. Communication among these planes is established by using well-defined Application Programmable Interfaces (APIs). These interfaces are divided as; Southbound APIs, Northbound APIs and in case of distributed controllers, East/Westbound APIs. Control plane and data plane communicates through Southbound API. It provides the information of data plane devices to control plane and push instructions/rules from control plane to data plane devices. Control plane and management plane uses Northbound API to provide programmability in SDN. Inter controller communication of SDN domains is established using Eastbound API, whereas Westbound API is responsible for legacy domain to SDN domain communication.

The southbound interface can be segregated into OpenFlow (OF)~\cite{OpenFlow}, OF dependent, and OF independent proposals. OF is the most commonly used SBI in research and commercial SDNs. Extensions of SBI into emerging technologies, such as sensor networks and Internet of Things, can also be treated as a special class of SBIs, due to their unique requirements. Similarly, northbound interfaces are classified in terms of portability, programmability, controller based, and intent based solutions. This work considers virtualization as a middle ware between different layers and the hardware. Hence, the interaction of APIs with different virtualization techniques requires separate classification. Eastbound interfaces are categorized in distributed and hierarchical architectures due to placement and communication of controllers, where as westbound interfaces usually use traditional Border Gateway protocols to bridge the gap between SDN and traditional networks. 

Prior to SDN, the concept of network programmability was studied from two aspects: Active Networking  \cite{active} and Programmable Networks \cite{PN} (A\&PN). Active Networking discusses the injection of intelligence in the network beyond conventional processing of packets. In active networks, nodes are capable of performing customized operations on packets passing through them. Whereas, Programmable Networks allows to control the behavior of network devices and flow control through software. This lays a clear foundation for separation between data and control plane. Since mid 90s and early 2000s these two concepts have combined to become Software Defined Networking paradigm. Neither SDN is the first nor the only solution that allows such kind of separation and programmability. The main reason for being wide acceptance of this paradigm is the rapid innovation in both planes (i.e. control plane and data plane) \cite{ForCES_Iqbal1}.

Open Networking Foundation (ONF) \cite{ONF} is leading in standardization of SDN and has support of more than hundred organizations. SDN has been extensively studied in academia as well as industry \cite{B4}, 
and has been implemented in different domains; e.g. Data Center (DC), Wireless Sensor Network (WSN), and Internet of Things (IoT) for management. Data Centers are ever-changing systems and have a large number of infrastructure assets, with complex topological configurations. Data Center management with traditional elements is very challenging because of their proprietary procedures rather than a unified process. Implementation of policies becomes very complex and time consuming. SDN makes DC management simple and straightforward using centralized programmable entities. It helps to apply a unified process on underlying devices using standardized southbound APIs which implement policies seamlessly. Similarly, WSN can benefit from SDN to resolve management, configuration, and optimal data transmission path challenges \cite{MANNA,WSNManagement,SDWNManagement}. By using Network Management System (NMS) at the management plane, the resource utilization and configuration from a single point becomes easy. Southbound Interfaces update the control plane for device changes and topological modifications. Internet of Things (IoT) which is more complex than WSNs, due to heterogeneity of devices, and connectivity to Internet, can also benefit from SDN.



\subsection{Major Contributions}
There are a number of studies which discusses implementation of SDN, controllers and APIs in general. However, there is no work, which survey in depth the SDN interface literature. In this article, we present a systematic survey and classification of different APIs. Main contributions include:
\begin{itemize}
\item Classification of each interface layer, (i.e. Southbound, Northbound and East/Westbound) into categories, based on their architectures and properties.
\item Survey in technical depth of each solution presented for each category.
\item Comparative analysis among APIs. As different interfaces have different properties and purpose, the comparison criteria is also designed accordingly.
\item Survey extended to interfaces defined for emerging SDN technologies for wireless sensor networks and Internet of Things.
\item Comparison of different SDN interfaces, which utilize (or benefit from) virtualization techniques sued in SDN.
\end{itemize}

In this paper we discuss a comprehensive detail of different APIs and groups them accordingly. We started from most commonly used southbound interface (i.e. OpenFlow) and classified various studies as dependent and independent on OpenFlow. Moreover, OpenFlow like solutions for WSN and IoT are investigated. Similarly, northbound interfaces are also classified into portability, programmability, controller based, and intent based solutions. SDN combined with virtualization is a new paradigm, and has brought a drastic growth in networks. Both southbound and northbound interfaces are involved in various virtualization schemes, which are grouped separately. To establish communication between SDN domains and with legacy networks are also classified. This paper also identifies a number of future research directions with respect to each plane and SDN component.
\subsection{Related Work}
Comprehensive study of SDN is a difficult task as it is a multi-dimensional field. However, it has been explored thoroughly with the aspects of its history, current state and future applications in a number of studies \cite{SDNSurvey1,SDNSurvey2,AdvSDN,KARAKUS2017279,MultiControllerSur}. In \cite{SDNSurvey1}, Masoudi et al. investigates data, control and application planes of SDN paradigm in details. Different simulation tools, debuggers, and testbeds for development of various aspects of SDN are also discussed. Gong et al. \cite{SDNSurvey2} describes different efforts on SDN architecture, component design, and applications. It also analyzes and provides a comparison of different applications in traditional and SDN environments. Recently, \cite{AdvSDN} is another attempt which provides a snapshot of current SDN state, a combination of benefits, challenges and opportunities, along with different simulation environments of SDN for testing purposes.  \par
A number of studies on control plane scalability and other issues (e.g. consistency, reliability etc.) of SDN architecture are presented in \cite{KARAKUS2017279,MultiControllerSur}. Karakus et. al \cite{KARAKUS2017279} highlights scalability problem of SDN architecture and categorizes this issue in different approaches as distributed (flat), hierarchical and hybrid designs. In \cite{MultiControllerSur}, Hu et al. presents different researches in multiple controller scenarios. This study is classified into four aspects; scalability, consistency, reliability, and load balancing. Trois et al. in \cite{ProgSurvey} explained multiple high-level languages and discussed some challenges. A number of programming languages with different functionalities to solve various issues are presented.  \par
Furthermore, researches about emerging technologies in SDN like WSN, IoT and Virtualization are discussed in \cite{SDWSN,SDSNSurvey,SDIoTSur,HyperSurvey,NFVSurvey}. Authors in \cite{SDWSN,SDSNSurvey} discuss different approaches for the infusion of SDN in WSNs to encounter the imperfections of WSNs. It is also envisioned that these two technologies can perform a key role in the looming of IoT. A synthesized view of IoT deployment in current state is described in \cite{SDIoTSur} and presents different SDN technologies from the aspects of IoT. In \cite{HyperSurvey}, a comprehensive survey on hypervisors is conducted where different hypervisors are categorized according to their architectures and execution platforms Whereas, Li et al. \cite{NFVSurvey} presents a relationship between NFV and SDN and highlights main challenges in Software Defined NFV architecture. \par

\subsection{Organization of Paper}
\begin{figure}
	\centering
  \includegraphics[width=0.95\linewidth]{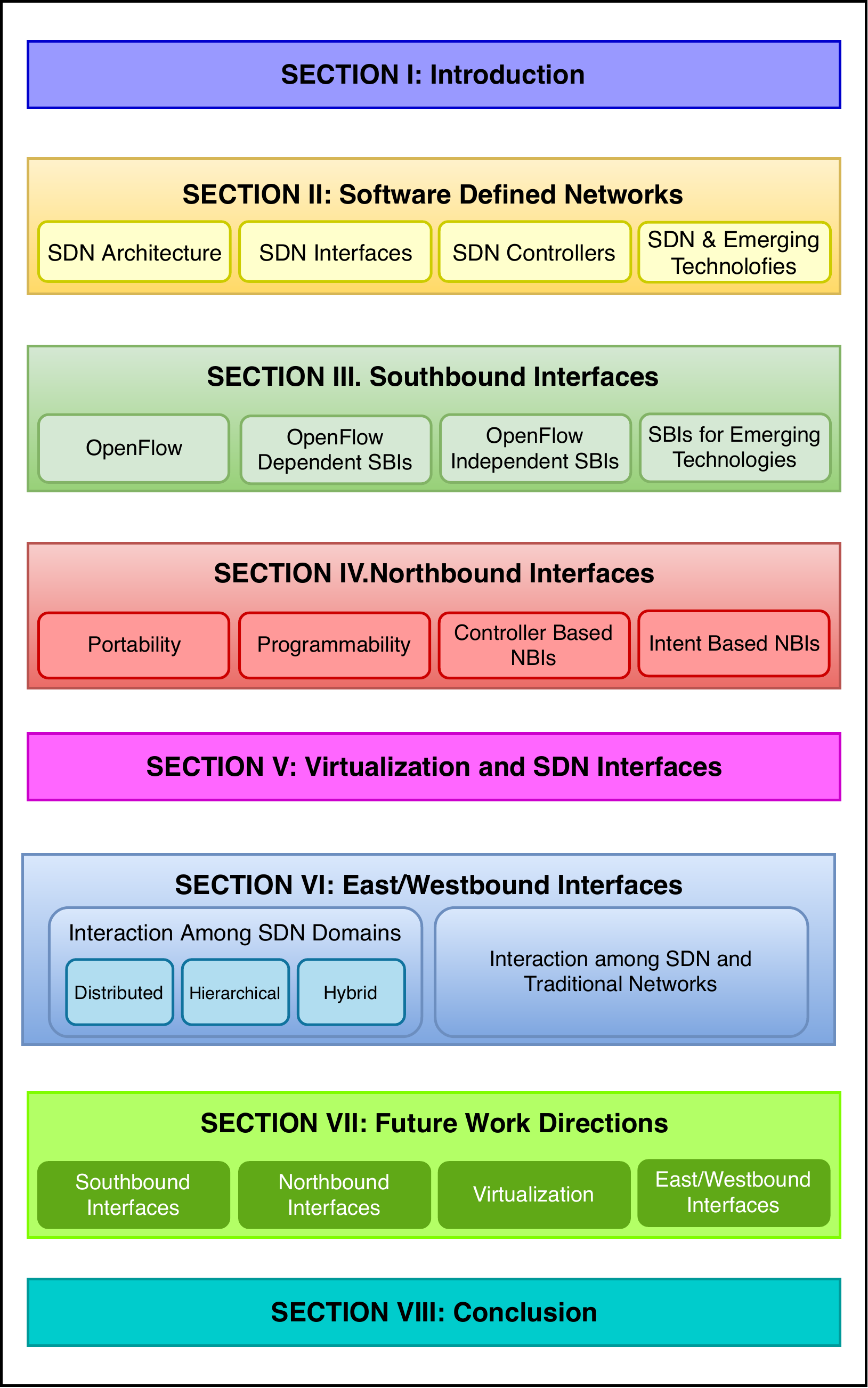}
  \caption{Overview and structure of this survey.}
  \label{fig:2}
\end{figure}

The organization of this paper is outlined in Fig. \ref{fig:2}, where Section II describes Background of SDN, its architecture, planes, and elements. Also, a brief introduction of emerging technologies (e.g. WSN, IoT, and NFV) is discussed. OpenFlow and other proposals for Southbound Interfaces, dependent and independent of OpenFlow are presented in Section III. Moreover, it presents southbound interfaces for WSN and IoT. Northbound interfaces are presented in Section IV with the aspects of portability and programmability etc. Virtualization in SDN is discussed in Section V along with the relationship of these interfaces and virtualization techniques. Communication among SDN domains and with legacy networks is presented in Section VI. Section VII outlines future directions, conclusion is given in Section VIII. 

\section{Software Defined Networks}
The main objective of this paper is to study the interfaces among different planes of SDN. Before delving into such details, we first give an overview of SDN architecture, its different components and their functionalities. We also discuss the use of SDN in emerging technologies which includes Wireless Sensor Networks (WSN), Internet of Things (IoT), and Network Function Virtualization (NFV).
\subsection{SDN Architecture}
In traditional networking, control functionalities and data forwarding elements are embedded in same device. Due to tightly coupled control and data planes, network control is distributed in these devices. It is hard to run applications on these elements due to vendor and language specifications. Moreover, management and configuration of such devices becomes difficult with increasing scalability. SDN paradigm breaks this tight coupling of control functions and forwarding elements, and provides a unified control for running different types of applications. Using SDN, networks can be divided into three planes; Data Plane, Control Plane and Management Plane as shown in Fig. \ref{fig:3}. 
\subsubsection{Data Plane}
Data Plane, also referred as Forwarding Plane, includes software or hardware switches and other physical devices. As the routing control is removed from the devices, hence SDN only uses programmable switches. These switches are capable of communicating with controller, usually using OpenFlow. They have three main components; Flow tables, secure channel, and OpenFlow protocol. Devices may have one or more flow tables. Secure channel connects it with the controller, and OpenFlow protocol provides communication with external controller. Flow table comprises of flow entries in the format of match, actions, and counters. For each packet, header matching is done and based on this matching a particular action is taken and counter is updated accordingly. Instructions are installed by control plane using southbound interfaces. Using these instructions, data plane devices can perform a number of actions which includes; forward to port, forward to controller, and drop. As soon as switch receives a packet, it first checks in the flow tables by matching its packet header. If it finds a flow entry against this matching, it takes action and forwards the packet to a particular port. Whereas, if it does not find any entry, packet is either dropped or forwarded to external controller through $Packet\_IN$ message. Based on the global view of network, controller replies as $Packet\_OUT$ message and installs flow entry for this packet. 
\subsubsection{Control Plane}
Control Plane or controller is the decoupled entity from data plane and has global view of whole network under its domain. It uses Network Operating System (NOS) to facilitate network management. It also acts as a strategic control point in SDN architecture, and works as a bridge between data plane and management plane. It manages flow control in the switches and elements using southbound interface. At the same time it uses northbound APIs to communicate with management plane for network applications. Based on its global view, it instructs data plane devices by installing flow entries and provides network state information to management plane for application development. SDN controller normally contains a set of modules that can perform different tasks. It gathers network statistics and performs inventory of network devices. To support more advance capabilities, extensions can be inserted. SDN controllers have very diversified properties. Almost all the controllers offer basic network services which includes; topology manager, stats manager, routing module, device manager, etc. Some controllers like NOX \cite{NOX}, POX \cite{POX} have centralized architecture and controllers like FloodLight \cite{FL}, OpenDaylight \cite{ODL} are distributed in nature. Most of the controllers support OpenFlow as southbound interface but OpenDaylight \cite{ODL} supports a wide range (i.e. OpenFlow, OvSDB, SNMP, NetConf) of southbound interfaces. A number of different controllers have been proposed in  \cite{NOX,POX,ODL,FL,Kandoo,DISCO}.
\subsubsection{Management Plane}
Management plane plays a vital role in SDN where network applications can be realized. SDN can be implemented to a wide range of networks, from home to enterprise and data center networks. This wide range of network environments leads to a variety of applications such as routing, policy enforcement, load balancing, and firewalls etc. As an example scenario of load balancing \cite{DLPO}, an application can take the appropriate actions to seamlessly distribute the traffic evenly among available multiple paths. Implementation of such a wide array of applications in traditional networks is very hard where control plane of each device needs to be configured independently. Using SDN, management plane provides a straightforward solution to implement these applications.
\subsection{SDN Interfaces}
\begin{figure}
  \centering
  \includegraphics[width=\linewidth]{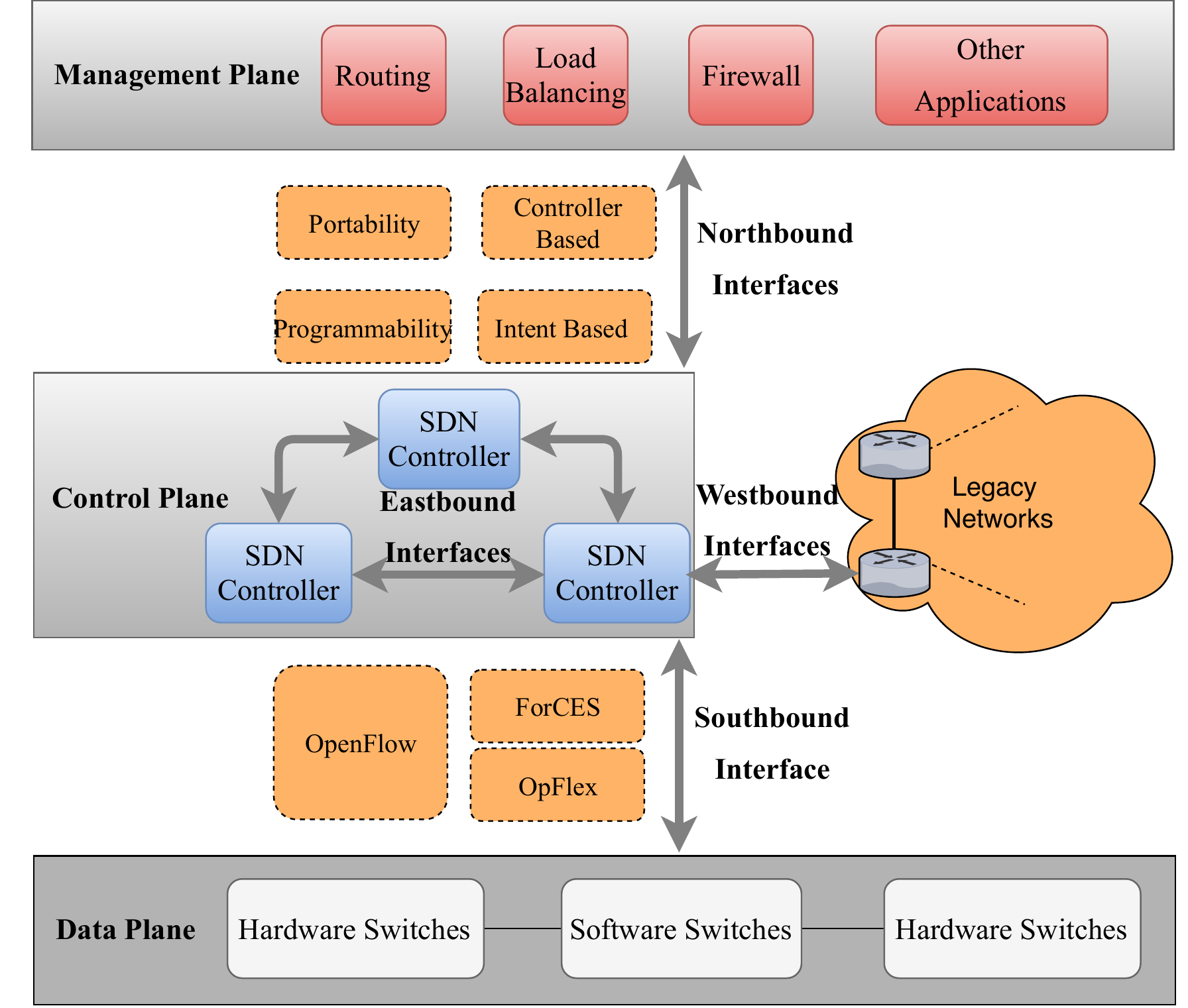}
  \caption{SDN interface placement and properties.}
  \label{fig:3}
\end{figure}

Application Programmable Interfaces (APIs) play a vital role in SDN and provide interaction among the planes. These APIs are architectural component of SDN and used to push configurations or information to forwarding elements or applications respectively. Fig~\ref{fig:3} shows the different interface APIs, and some of their properties in and SD network.
\subsubsection{Southbound API}
Southbound API (SBI) is an SDN enabler, which provides a communication protocol between control plane and data plane. This API is used to push configuration information and install flow entries in data plane. It also provides an abstraction of network device's functionality to the control plane. Major challenges of southbound interfaces are heterogeneity, vendor specific network elements, and language specifications. There is a wide range of traditional network elements which creates heterogeneity issues. Every vendor has its own architecture for switching fabric and supports different languages. Southbound interface in SDN resolves these issues by providing an open and standardized interface. There are a number of examples for Southbound APIs but OpenFlow \cite{OpenFlow} is considered as a standard in SDN. Section III presents a detailed review of SBIs.

\subsubsection{Northbound API}
The numerous benefits of SDN are fruitless, if applications can not benefit. SDN adoption depends on its ability to support a wide range of applications. Northbound APIs (NBIs) play an integral role for application developers and provides a common interface between controller and management plane. It helps to provide the information of underlying devices for application development which makes SDN control easy and dynamic. Unlike southbound interface, northbound interface has seen less standardization efforts \cite{6994333} because it is a software ecosystem. A wide range of NBIs are offered by current controllers and programming languages. In this paper, we discuss them from different aspects in Section IV and V.
\subsubsection{East/Westbound API}
Centralized control over the network is the key feature of SDN, but only a limited number of switches can be handled by a single controller. Due to an exponential increase in the network devices and for large scale networks, distributed controllers become a requirement. In such a distribution, every controller has its own domain with underlying forwarding devices. Controllers need to share information of their respective domains  for a consistent global view of the entire network. Eastbound APIs are used to import and export information among distributed controllers. Some examples of these interfaces are \cite{SDNi,AMQP,RabbitMQ}. On the other hand, Westbound APIs enable the communication between legacy network devices (routers etc.) with the controllers. Some example solutions are discussed in \cite{RouteFlow,SDNIP,BTSDN}. Detailed discussion and review is given in Section VI.
\subsection{SDN Controllers}
A controller is the fundamental component of SDN control plane. One of the key role of a controller is to manage the traffic in underlying network elements by using a set of instructions, referred as flows. There is a wide range of controllers in SDN, however we present only the most common controllers, their architecture, and interfaces in Table \ref{tab:1a}.

\begin{table}[!t]
	\centering
 	\caption{Summary of SDN Controllers}
 	\label{tab:1a}%
 	\tiny
	\setlength\tabcolsep{2.5pt}
	\begin{tabularx}{\linewidth}{|>{\RaggedRight\hsize=\hsize}X|>{\RaggedRight\hsize=\hsize}X|>{\RaggedRight\hsize=0.8\hsize}X|>{\RaggedRight\hsize=1.2\hsize}X|>{\RaggedRight\hsize=1.2\hsize}X|>{\RaggedRight\hsize=0.8\hsize}X|}
		\hline
		\textbf{Controller} & \textbf{Programming Language} & \textbf{Architecture}  & \textbf{SBIs}  & \textbf{NBIs}  & \textbf{EWBIs}   \\ \hline
		
		NOX \cite{NOX} & C++  & Centralized  & OpenFlow 1.0  & ad-hoc  & -  \\ \hline
		POX \cite{POX} & Python  & Centralized  & OpenFlow 1.0  & ad-hoc  & -  \\ \hline
		Floodlight \cite{FL} & Java  & Centralized  & OpenFlow 1.0-1.3  & REST, Java RPC, Quantem  & -  \\ \hline
		OpenDaylight \cite{ODL} & Java  & Distributed  & OpenFlow 1.0-1.3, OvSDB, SNMP  & REST, RESTCONF, XMPP, NETCONF  & SDNi  \\ \hline
		Kandoo \cite{Kandoo} & C, C++, Python  & Distributed  & OpenFlow 1.0-1.2  & Java RPC  & Messaging Channel  \\ \hline
		DISCO \cite{DISCO} & Java  & Distributed  & OpenFlow 1.0  & REST  & AMQP  \\ \hline
		Beacon \cite{beacon} & Java  & Centralized  & OpenFlow 1.0  & ad-hoc  & -  \\ \hline
		ONOS \cite{ONOS} & Java  & Distributed  & OpenFlow 1.0-1.3  & REST, Neutron  & Raft  \\ \hline
		Onix \cite{Onix} & C++  & Distributed  & OpenFlow 1.0, OvSDB  & Onix API  & Zookeeper  \\ \hline
		PANE \cite{PANE} & Haskell  & Distributed  & OpenFlow 1.0  & PANE API  & Zookeeper  \\ \hline
		HyperFlow \cite{Hyperflow} & C++  & Distributed  & OpenFlow 1.0  & -  & WheelFS  \\ \hline
	\end{tabularx}
\end{table}

Features of each controller may differ from one another, but core and essential functionalities of all the controllers are similar, for example, topology information, statistics, notifications, and device management. To perform these tasks, every controller uses a southbound interface such as OpenFlow.  However, some of the controllers offer a wide range of southbound interfaces (e.g. OpenDaylight). In order to run various applications, every controller offers a northbound interface which lacks in standardization as compared to southbound interface. \par
Although controllers can be categorized with many aspects, however one of the key aspect is architecture where controllers can be classified into either centralized or distributed. In centralized controllers, a single entity is responsible for managing all the network devices. Whereas, in distributed controllers a number of entities cooperate with each other to manage the underlying elements. East/westbound interfaces are the key enabler for distributed environments.

\subsection{SDN for Emerging Technologies}
SDN has moved from local networks and data centers to new domains which can be beneficial as well as challenging. In this section we will give a brief introduction of use of SDN in emerging technologies to solve the challenges faced in these domains. 
\subsubsection{Wireless Sensor Networks}  
Since 2000s, Micro-Electro-Mechanical Systems (MEMS) have matured to a great extent, where they could be used to build large scale wireless networks \cite{WSN1}. These low power devices sense information from environment, process and forward it to remote locations \cite{WSN2}. WSNs have been deployed in widespread applications from civil to military, and from environmental to health care.

One of the major challenge in WSN is that it has devices with limited power resources and energy capabilities. SDN, on the other hand, separates control and data planes and shifts most of the energy intensive functions, such as routing, from underlying device to centralized controller. Another issue is that, it provides application specific solutions and to solve this issue, a significant research has been devoted for programmable WSNs \cite{WSN3,WSN4}. But without availability of operating system, programmers need to focus on intensive low level details. SDN provides a centralized controller working as an operating system. It also provides high level programming abstractions, simplicity and evolvability which provides simplified solutions and easy configuration of network devices. As a result, higher efficiency of network equipment can be achieved.
  
Another limitation in WSNs is rigidity in policy changes and is being solved by using local algorithms. Because of the low level programming it is very hard to change the policies in WSNs which is an ever-changing process. This issue can also be resolved by using management plane of SDN through northbound interface which provides flexibility in terms of policy making and implementation on it. There are still many areas of improvement in WSN where SDN can be a potential solution, especially radio resource management etc.

\subsubsection{Internet of Things} 
Internet of Things (IoT) \cite{IoT_Iqbal} is a network of physical devices, sensors, home appliances, and other electronic devices that can connect to the Internet and transfer data freely. The main concept behind IoT is to connect devices (virtually or physically) with a unique ID seamlessly. Some major applications of IoT are: smart homes, smart cities, autonomous vehicular networks etc. The widespread growth of IoT involves major challenges of management as well as communication mechanisms between heterogeneous objects~\cite{IoT3}.

According to \cite{IoT4}, IoT has a potential value of \$14 trillion over next 10 years. To handle such a large number of devices, flexible network management is required. Traditional networks use switches/routers. These devices are programmed with complex rules which makes these devices incapable to meet the requirements of IoT in real time. By using centralized SDN controller, an adequate mechanism can be implemented on forwarding devices.

The heterogeneous architecture of IoT also poses a challenge in optimizing the flow of information. Distributed controllers in SDN can resolve this challenge, where a controller communicates with other SDN controllers to exchange network-wide information. It also supports load sharing and load balancing among several different controllers. 

\subsection{Network Function Virtualization}
Network Function Virtualization (NFV) offers new ways to deploy, design, and manage networking devices. It separates network functionality, such as firewalls, Network Address Translation (NAT), and Domain Name Service (DNS), from hardware devices, so that it can be run remotely. European Telecommunication Standard Institute (ETSI) introduced Network Function Virtualization which is implemented as Virtual Network Function (VNF) \cite{VNF1}. NFV is highly complementary to SDN but not dependent on it.

SDN controller functions can be deployed as virtual functions also. OpenFlow switches can be controlled using NFV software. Emergence of these two technologies allows replacement of dedicated and expensive hardware equipment by software. Multi-tenancy requirements of cloud also uses NFV to support SDN. If a tenant does not require a full fledge controller, NFV can help SDN by virtualizing SDN controller so that different tenants can share a single controller. On the other hand, to achieve optimized traffic engineering between different VNFs, SDN can provide programmable network connectivity.
\begin{figure}[!t]
	\centering
  	\includegraphics[width=0.94\linewidth]{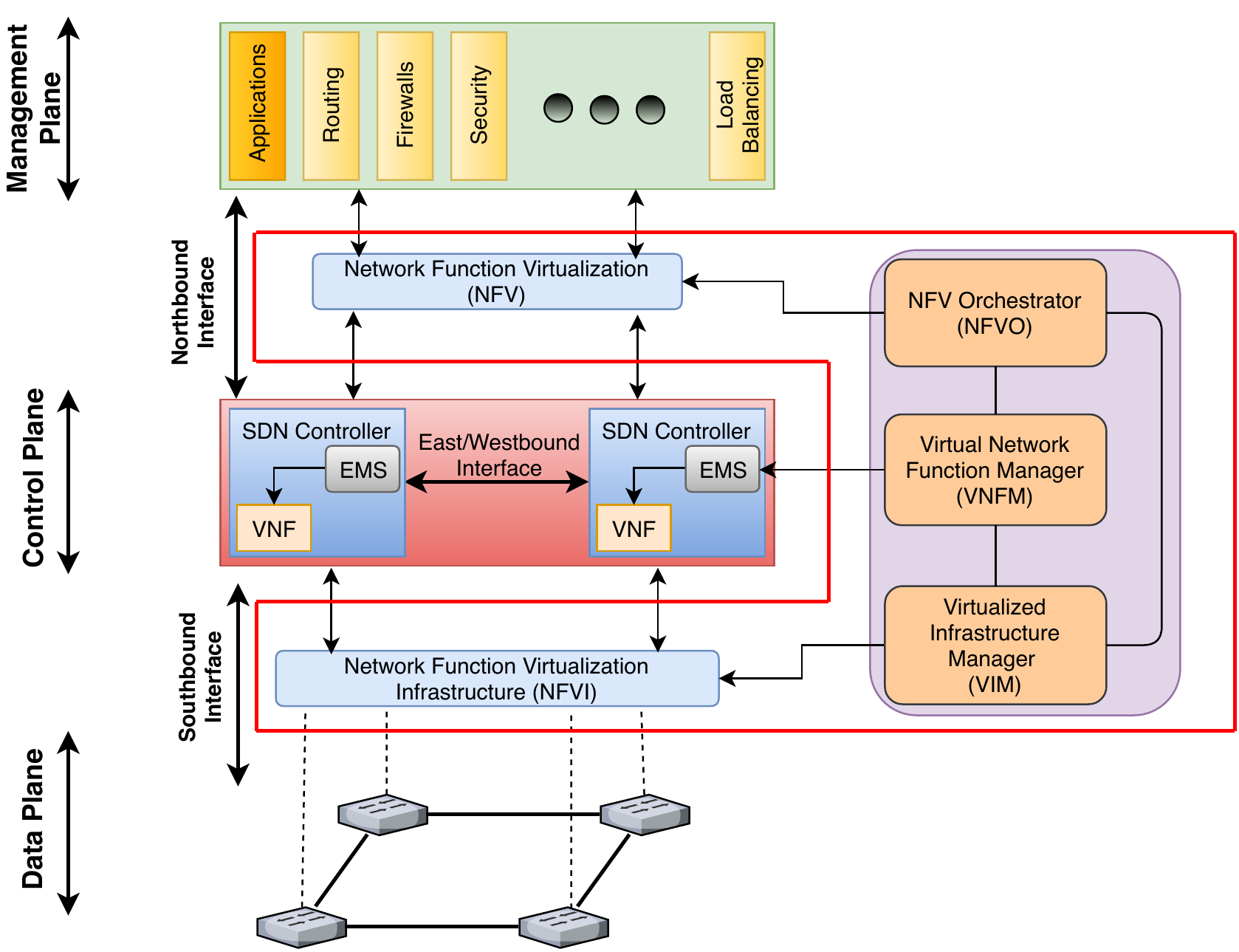}
  	\caption{Basic SDN and NFV interface abstraction.}
  	\label{fig:4}
\end{figure}

Control plane in SDN is encapsulated by NFV as shown in Fig. \ref{fig:4}. Virtual Infrastructure Manager (VIM) plays an integral role in NFV framework. It normally controls and manages NFV infrastructure storage, computing, and network resources. It also keeps a mapping of allocation of virtual resources to physical resources, and manages virtual networks, links, and ports. Virtual Network Function Manager (VNFM) is capable of handling multiple Virtual Network Functions (VNFs) by using Element Management System (EMS). To ensure an adequate availability of computing, storage and network resources, NFVO can either work directly with VIM or through VNFM. NFV can be utilized in two ways. One is virtualization of network resources by making slices (e.g. Hypervisors) where southbound interface is involved. The other is to control these slices (e.g Applications based Virtualization) which involves northbound interface. Both of these interfaces (Northbound and Southbound) fall under NFV.

\section{Southbound Interfaces (SBIs) in SDN} \label{sec2}
Software Defined Networking separates the network control and forwarding functions. With the help of southbound interfaces, forwarding function is kept on the device whereas, network control is shifted to an external controller. Southbound APIs enable link between the data plane and the control plane. It is very important that this link remains available and secure, otherwise the forwarding elements can not function. 

Main objective of this interface is to push notifications given by controller, to data plane devices, and provide information of these devices to controller. It allows the discovery of network topology, define network flows, and implement requests sent by management plane. 
 
Some of commonly known southbound interfaces are OpenFlow (OF) \cite{OpenFlow}, FORwarding \& Control Element Separation (ForCES) \cite{ForCES_Iqbal1}, Open virtual Switch Database (OvSDB) \cite{OVSDB_Iqbal}, Protocol Oblivious Forwarding (POF) \cite{POF_Iqbal}, OpFlex \cite{OpFlex}, OpenState \cite{OpenState_Iqbal}, etc.
  
 \begin{figure}[!t]
 	\centering
 	\includegraphics[width=\linewidth]{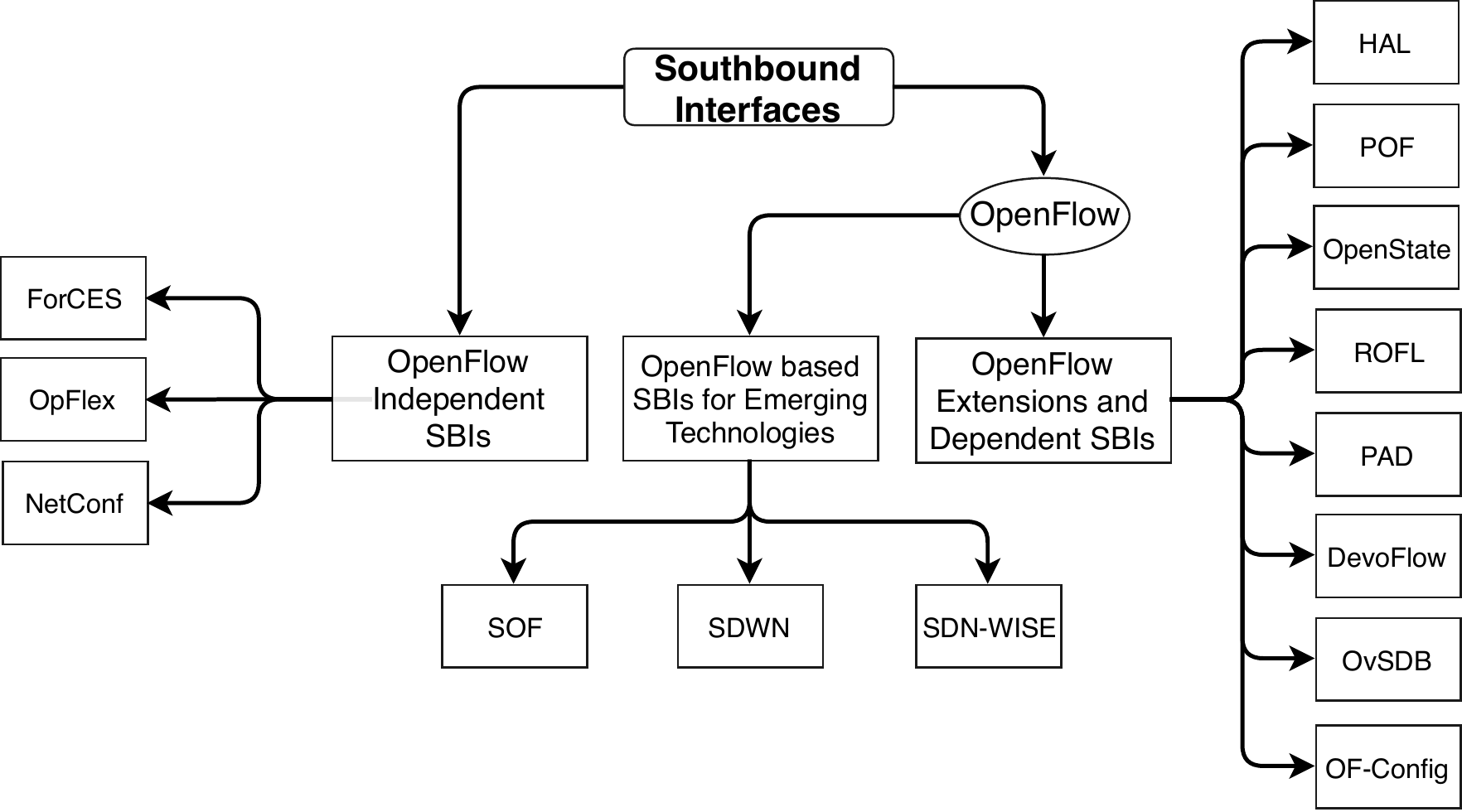}
 	\caption{Classification of OpenFlow dependent and independent SBI proposals.}
 	\label{fig:5}
\end{figure}

Fig. \ref{fig:5} classifies different southbound interfaces proposals. OpenFlow \cite{OpenFlow} is most commonly used API and considered as a standard southbound interface. Most of the proposals are either extensions or somehow dependent upon OpenFlow. Two proposals for southbound interfaces (i.e. OvSDB and OF-Config) work as OpenFlow companions and help it to provide configuration capabilities. Whereas, proposals like, ForCES, OpFlex, and NetConf, are totally independent of OpenFlow. Proposals like Sensor OpenFlow (SOF) \cite{SOF}, Software Defined Wireless Networks (SDWN) \cite{SDWN}, and SDN for WIreless SEnsors (SDN-WISE) \cite{SDNWISE} are southbound interfaces defined specifically for emerging technologies. In the following subsections, we first discuss OpenFlow, followed by its dependent and independent SBI literature. We also discuss the SBIs for sensor networks and Internet of Things.

\subsection{OpenFlow}
\subsubsection{OpenFlow Evolution}
OpenFlow \cite{OpenFlow} is a standardized and most commonly used southbound interface. It was designed particularly for SDN to provide communication between controller and forwarding elements. OpenFlow has evolved from version 1.0 with only 12 fixed matching fields and a single flow table, to version 1.5 with 41 matching fields and a number of new functionalities. Table \ref{tab:1} shows the evolution of OpenFlow and addition of features to it. Here we list the major features, not including minor optimizations and performance improvements.

\begin{table*}[]
	\centering
	\caption{OpenFlow Version Timeline and Major Changes}
	\begin{tabularx}{\textwidth}{|>{\hsize=0.8\hsize}X|>{\centering\hsize=0.5\hsize}X|>{\hsize=1.5\hsize}X|>{\hsize=1.2\hsize}X|}
		\hline
		\textbf{Version}	&\textbf{Year}	&\textbf{Major Features}	&\textbf{Reasons of Extension}  \\ \hline
		\multirow{2}{*}{OpenFlow 1.0~\cite{V0}} & \multirow{2}{*}{2009} & Single Table	&-  \\ \cline{3-4} 
											&	&\multicolumn{1}{l|}{Fixed Matching Fileds} &-  \\ \hline
		
		\multirow{3}{*}{OpenFlow 1.1~\cite{V1}} & \multirow{3}{*}{2011} & Multiple Tables	& Avoids Flow Entry Explosion \\ \cline{3-4} &
		     & Group Tables		&Action Set to a Group of Tables \\ \cline{3-4}
		     &	& VLAN and MPLS Support	& - \\ \hline
		     
		\multirow{3}{*}{OpenFlow 1.2~\cite{V2}} & \multirow{3}{*}{2011}  & OpenFlow eXtensible Match Using TLV Structure &Increased Matching Flexibility\\ \cline{3-4}
			&	& \multicolumn{1}{l|}{IPv6 Support} &-\\ \cline{3-4}
			&	& Controller Role Exchange	& Controller Scalability \\ \hline
			
		\multirow{2}{*}{OpenFlow 1.3~\cite{V3}} & \multirow{2}{*}{2012}	& Meter Table	& Add Quality of Service\\ \cline{3-4} 
		    &   & \multicolumn{1}{l|}{Table-miss Entry} & Provides Flexibility  \\ \hline
		    
		\multirow{2}{*}{OpenFlow 1.4~\cite{V4}} & \multirow{2}{*}{2013} & Synchronized Table	& Enhances Table Scalability\\ \cline{3-4} 
		    &	& \multicolumn{1}{l|}{Bundle Supports Group Modifications} & Enhances Switch Synchronization  \\ \hline
		    
		\multirow{2}{*}{OpenFlow 1.5~\cite{V5}} & \multirow{2}{*}{2015}	& Egress Table	& Allows processing on output ports\\ \cline{3-4} 
		    &   & \multicolumn{1}{l|}{Scheduled Bundle}	& Extends Switch Synchronization Further  \\ \hline
	\end{tabularx}
	\label{tab:1}
\end{table*}

OpenFlow version 1.0 \cite{V0} had only one flow table with three components: Header fields, counters, and actions. Moreover, it did not have much flexibility due to fixed matching fields. Switches using this version could not perform more than one operation during packet forwarding because of single flow table which directly effected usability and scalability of OpenFlow.

To overcome these problems, OpenFlow version~1.1~\cite{V1} introduced multiple tables where \textit{packet process pipeline} was used. The difficulty in implementing a pipeline was resolved by Forwarding Abstraction Work Group (FAWG) \cite{FAWG} which proposed Negotiable Datapath Model (NDM). NDM is an abstraction of forwarding model supported by network devices, which is normally expressed in JavaScript Object Notation (JSON). Another new feature in this version was Group Table, with group entries which are divided in four types: All, Select, Indirect, and Fast Fail-over. Header field was renamed as Match field due to the fact that header field did not exactly match the headers.

OpenFlow version~1.1 used fixed length structure for match fields which in version~1.2~\cite{V2} was changed to Type-Length-Value (TLV) structure to provide more flexibility. This is referred to as OpenFlow eXtensible Match (OXM). In addition, it added support for IPv6 based on OXM. During this time single controller was considered as single point of failure and posed a threat for failover and load balancing. Thus, OpenFlow~1.2 introduced controller role-change mechanism by which multiple controllers could exist as master/slaves or equals, which resolved failover and availability problems. 

One of the essential features of computer networking is Quality of Service (QoS) and to enhance that, OpenFlow version~1.3~\cite{V3} introduced Meter Table having multiple Meter Entries known as Meter Identifiers. It also extended the flow table with table-miss entry. In previous version a packet used to be either forwarded to a particular port or dropped, but table-miss entry provided more flexibility in OpenFlow by sending packet to the controller.

Synchronized table was introduced in OpenFlow version~1.4~\cite{V4} for the scalability of flow tables, and also enabled unidirectional and bidirectional tables. If table synchronization is bidirectional then any changes done by the controller will be reflected on the source table which is effective when switches are doing multiple lookups upon same lookup data. MAC forwarding table and MAC learning table of an Ethernet switch is an example when both of these tables lookup on same MAC address.  Another feature included in this version was Bundle which is used to do modifications on a group of switches.

OpenFlow version 1.5 \cite{V5} further extended bundles as Extended Bundle which strengthen the synchronization among multiple switches. In addition, it introduced Egress Table which allows packet matching based on its output port unlike previous versions which only uses to match and process ingress packets. 

\begin{figure}[!t]
	\includegraphics[width=\linewidth]{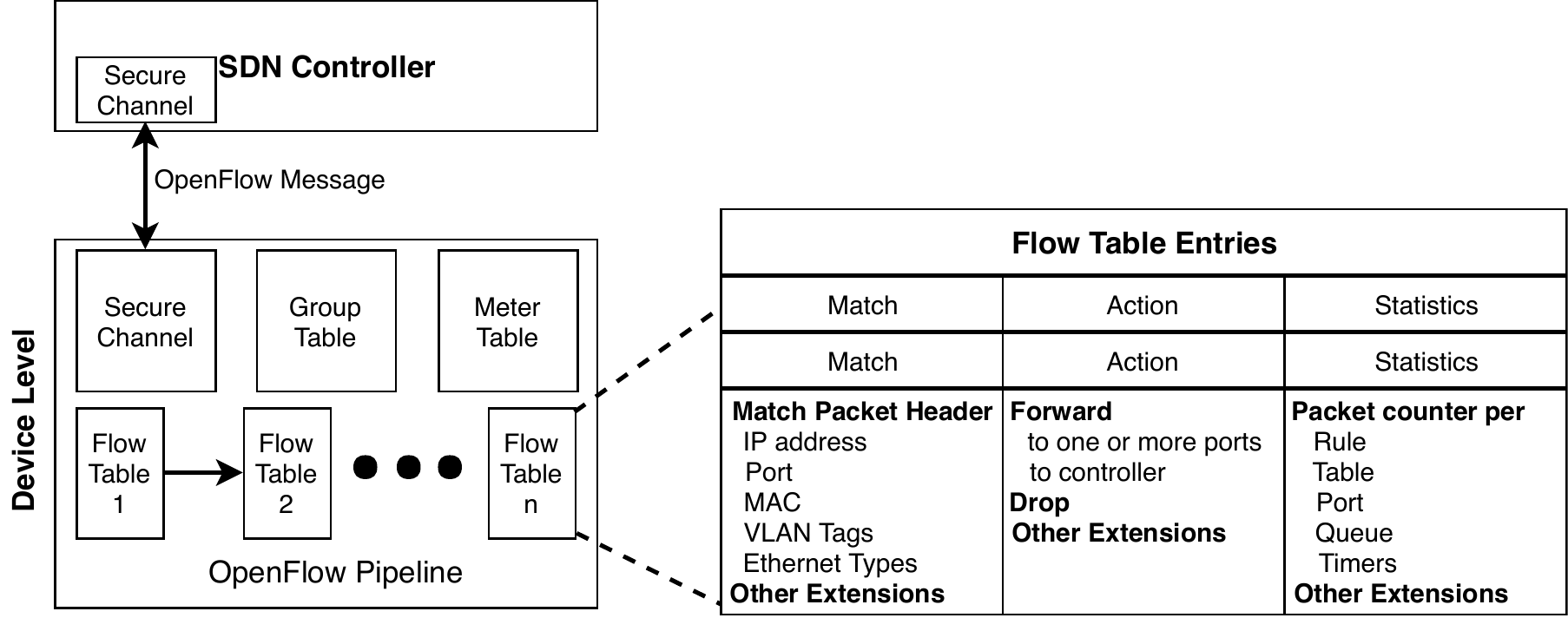}
	\caption{OpenFlow structure.}
	\label{fig:6}
\end{figure}

\subsubsection{OpenFlow Architecture}
OpenFlow enabled devices must have three main components; Flow Table, Secure Channel, and OpenFlow protocol. Devices may have one or more flow table, secure channel connects it with the controller and a protocol provides communication with external controller as shown in Fig. \ref{fig:6}. OpenFlow pipeline has a number of flow tables along with group table and meter table. Tables in OpenFlow enabled switches have flow entries in the format of match, actions, and statistics. For each packet, header matching is done which includes; source and destination IP, source and destination port, source and destination MAC, along with VLAN tags, and Ethernet types. Flow tables are normally numbered and starts from 0 and packet processing pipeline always starts from this table.

Based on this matching a particular action is taken to forward packet on one or more ports. If no match is found, then it is forwarded to controller using Packet\_IN message. This message contains the information of ingress port, packet header, and Buffer\_ID where the packet is stored. To respond \emph{Packet\_IN} message, controller sends a \emph{Packet\_OUT} message. This message contains \emph{Buffer\_ID} of corresponding \emph{Packet\_IN} message and actions to perform (e.g. Forward to a particular port, drop etc.). To handle the subsequent packets of same flow, controller sends a \emph{Flow\_Mod} message to switch with the instruction to insert rules into flow table. Given rules are matched for the subsequent packets of same flow and an action is taken at line speed. Meanwhile, counter is updated accordingly and statistics are generated per rule, per table, per port, per queue or per timer.

\begin{table*}[!t]
  \centering
  \caption{Summary of OpenFlow Dependent SBI Proposals}

	\setlength\tabcolsep{3pt}
	\begin{tabularx}{\linewidth}{|>{\RaggedRight\hsize=0.9\hsize}X|>{\RaggedRight\hsize=1.3\hsize}X|>{\RaggedRight\hsize=1.1\hsize}X|>{\RaggedRight\hsize=1.3\hsize}X|>{\RaggedRight\hsize=0.7\hsize}X|>{\RaggedRight\hsize=0.7\hsize}X|}
	\hline
    \textbf{Literature}	&\textbf{Objective}	&\textbf{Solution}	&\textbf{Benefits}	&\textbf{Network Type}	&\textbf{Resolved by OpenFlow?}\\\hline
    
    POF~\cite{POF_Iqbal} &
      Remove dependency on protocol specific configuration &
      Flow Instruction Set (FIS)  &
      Reduce network cost by using commodity forwarding elements &
      Not Mentioned &
      No
  \\ \hline
    OpenState~\cite{OpenState_Iqbal} &
      Reduce Overhead between switch and controller &
      Making devices as eXtensible Finite State Machines (XFSM) &
      Reduce Interaction between controller and forwarding elements &
      Not Mentioned &
      Partially Solved 
  \\ \hline
    ROFL~\cite{ROFL_Iqbal} &
      Development of OF enabled applications and support new OF versions and extensions &
      eXtensible DataPath daemon (XDPd) &
      Supports multiple versions of OF simultaneously  &
      Not Mentioned &
      \multicolumn{1}{l|}{No}
 \\ \hline
    HAL~\cite{HAL_Iqbal} &
      To realize OF functionality in legacy network devices &
      Cross Hardware Platform Layer and Hardware Specific Layer &
      Legacy Network devices to communicate with SDN domains  &
      Data Center Networks &
      Yes                 
  \\ \hline
    PAD~\cite{PAD_Iqbal}  & Expose switch capabilities and better utilization of network devices   & Generic Byte Operation & Works for devices which are not working on packets e.g. Optical Switches & 
      Not Mentioned &
      \multicolumn{1}{l|}{Yes}  \\ \hline

    DevoFlow~\cite{DevoFlow} &
      Control traffic overhead  &
      Giving control to switch and controller only manages targeted flows &
      Reduce Interaction between controller and forwarding elements but major changes required in switch &
      Data Center Networks &
      Partially Solved  \\ \hline
        OvSDB~\cite{OvSDB}\newline \emph{(For Configuration)}  & Enhances configuration capabilities of OpenFlow  & Uses database server and switch daemon & Provides better configuration & 
      Hybrid Networks &
      \multicolumn{1}{l|}{-}  \\ \hline
        OF-Config~\cite{ofconfig}\newline \emph{(For Configuration)} & Remote configuration of OpenFlow switches  & Uses configuration points & Provides flexibility for configuration in OpenFlow switches & 
      Hybrid Networks &
      \multicolumn{1}{l|}{-}  \\ \hline
\end{tabularx}
  \label{tab:2}%
\end{table*}

\subsection{OpenFlow Dependent SBI Proposals}
This sub-section describes the Southbound proposals which are based on OpenFlow and attempt to enhance its existing features or in its newer versions. Some of these issues have been addressed by OpenFlow fully, some are resolved partially, and few of the short comings are still present in OpenFlow protocol. Table \ref{tab:2} represents such solutions along with their objectives.

Enabling legacy network devices to become OpenFlow compliant is a complicated task. Hardware Abstraction Layer (HAL) \cite{HAL_Iqbal} attempts to resolve this issue. It decouples hardware-specific control and management logic from the network node abstraction, which hides device complexity and vendor specific features. HAL achieved this decoupling by introducing two sub-layers; Cross Hardware Platform Layer (CHPL) and Hardware Specific Layer (HSL). CHPL covers node abstraction, virtualization, and configuration mechanisms, whereas HSL is responsible for discovering the particular hardware platform and perform all required configuration using Hardware Specific Module (HSM). Furthermore, every network element in this environment has its own protocol for communication purposes, controls, and management of underlying system. HSL in HAL is used to hide this complexity and heterogeneity. Depending on the type of network devices, communication between these two layers is done by using Abstract Forwarding API and Hardware Pipeline API. Another interesting feature of HAL is to handle multiple versions of OpenFlow.

OpenFlow has underwent many changes since its initial versions and pace of development has required other third party hardware and software in data and control plane to make massive changes to their solutions. OpenFlow has detailed specification documents for each version but to create new libraries for each and every platform is time consuming. Revised OpenFlow Library (ROFL) \cite{ROFL_Iqbal} resolved this problem and provided a clean and easy to use API which hides the details of respective protocol versions (i.e. 1.0, 1.2 and 1.3), and simplifies application development. ROFL uses eXtensible Datapath daemon (xDPd) which is a framework for developing SDN datapath elements. It uses three major libraries, ROFL-common, ROFL-pipeline, and ROFL-HAL. ROFL-common is used to provide the basic support of OpenFlow protocol which is comprised of protocol parsers and message mangling. ROFL-pipeline is employed as data model whereas ROFL-HAL is implemented as an interface.

DevoFlow \cite{DevoFlow} addressed the overhead created by the OpenFlow, due to full control and visibility of all flows through the software controller. It claims that ratio of control plane to data plane is four orders of magnitude and less than its aggregate forwarding rate. Devoflow attempts at resolving this problem by devolving control of most flows back to switches while controller maintains control over targeted and significant flows only. In this way switch-controller interactions and Ternary Content-Addressable Memory (TCAM) entries may reduce overhead. Another target is to provide the aggregated flow statistics to maintain enough visibility of network, but this approach required major modifications in switch design which is a costly solution.

The same problem was addressed by OpenState \cite{OpenState_Iqbal} where authors argue that all control should not be given to centralized controller and making switch stateless is a compromise rather than choice which causes extra communication between devices and controller. It also suggests that the programmers can deploy states in the device rather than using an external controller. OpenState abstraction relies on Extended Finite State Machine (XFSM) that allow the implementation of several stateful tasks inside forwarding devices. It uses XFSM as an extension of OpenFlow match-action phenomenon and allows to implement several stateful tasks inside the forwarding element without introducing the overhead of controller. All the tasks, which involve local states, such as port knocking and MAC learning, can be executed directly in network elements without any overhead of control plane communication or processing delay.
 	
Another problem in OpenFlow reported by POF \cite{POF_Iqbal} is that it is reactive rather than proactive, and data plane needs to be protocol aware. Due to that reason data plane devices need to understand packet header in a specific format to extract keys and execute packet processing, which again causes overhead. Moreover, data plane is almost stateless and can not perform any action without involvement of controller, which means data and control plane are not properly decoupled hence leading to problems like: hindrance in innovation, reducing programmability potential, and causing complexity for large scale networks. To resolve all these problems POF proposed Flow Instruction Set (FIS) which makes forwarding elements as white boxes, protocol oblivious and ensures its elegance and simplicity. FIS is a protocol independent set of instructions which helps to compose network services from the control plane. At the same times it helps in completely decoupling control and data plane so that both of these planes can evolve independently.

In Programming Abstraction Datapath (PAD) \cite{PAD_Iqbal} authors exposes switch capabilities for programmability and provides a southbound API for other types of devices such as Optical Switches. It provides generic programming of forwarding devices by using byte operations which define protocol headers and functions. A packet received at ingress port using PAD is bound with metadata and processed through a search engine, which is a functional component of PAD. As a result of this search, a function name is added which will be executed on this packet in execution engine. Finally packet is forwarded to egress port for transmission. PAD is applicable to optical flows, where forwarding functions do not contain packet information but other instructions.

{\textbf{Configuration:}} There are two solutions which provide configuration in OpenFlow: OvSDB and OF-Config. These protocols form a relationship between controller and switches. OpenFlow determines route of the packet but it does not provide the management and configuration which is necessary to assign IP addresses or port allocation. In traditional networks, vendors normally use different configuration and management methods which either depend on protocols like SNMP or use command line interface. SDN provides holistic view of every component of network to engineers.

OvSDB \cite{OvSDB} is designed to be used as virtual switch to forward traffic between different virtual environments. It is an open source switch, hence open to programmatic extensions and control using OpenFlow, and based on client and server implementation. Open vSwitch is a complementary protocol to OpenFlow. Inside a virtual switch, there is ovsdb-server, ovesdb-daemon and optionally a forwarding path. A virtual switch uses OpenFlow as interface to communicate with control and management cluster. Furthermore, it allows creation of multiple virtual switch instances, set Quality of Service (QoS) policies on interfaces, and collect stats. Managers can specify number of virtual bridges by using OvSDB which allows to create, configure, and delete ports.

OpenFlow Configuration protocol (OF-Config) \cite{ofconfig} has a special set of rules to define the mechanism for controllers to access and modify the configuration data on OpenFlow switches. It works as a companion of OpenFlow protocol and allows the remote configuration of OpenFlow switches. Major difference between OpenFlow and OF-Config is that OpenFlow modifies match-action rules which effects flows in OpenFlow switch datapath. Whereas, OF-Config remotely configure multiple OpenFlow datapaths on a physical and virtual platform.

{\bfseries{Conclusion:}} Most of the solutions based on (or extensions of) OpenFlow address the short coming in it. Some of these solutions have already been adopted by OpenFlow, whereas some of these are still open challenges. One of the major problem in SDN is to accommodate traditional network devices. This has attracted a lot of research attention. OpenFlow version~1.3 \cite{V3} tried to resolve this issue by introducing OpenFlow hybrid, which supports both OpenFlow operations as well as traditional Ethernet switching. Similarly, OpenFlow version 1.4 \cite{V4} has added support to optical switches which is discussed in PAD. To reduce the overhead between controller and data plane devices OpenFlow proposed Stats-Trigger in version 1.5 \cite{V5} which solved the problem partially. However, issues like support of multiple versions of OpenFlow and protocol dependency are still open research challenges.
 
\subsection{OpenFlow Independent SBI Proposals}
In this sub-section southbound API proposals which are independent of OpenFlow or are parallel proposals have been discussed.

\begin{figure}[!b]
	\centering
  \includegraphics[width=1\linewidth]{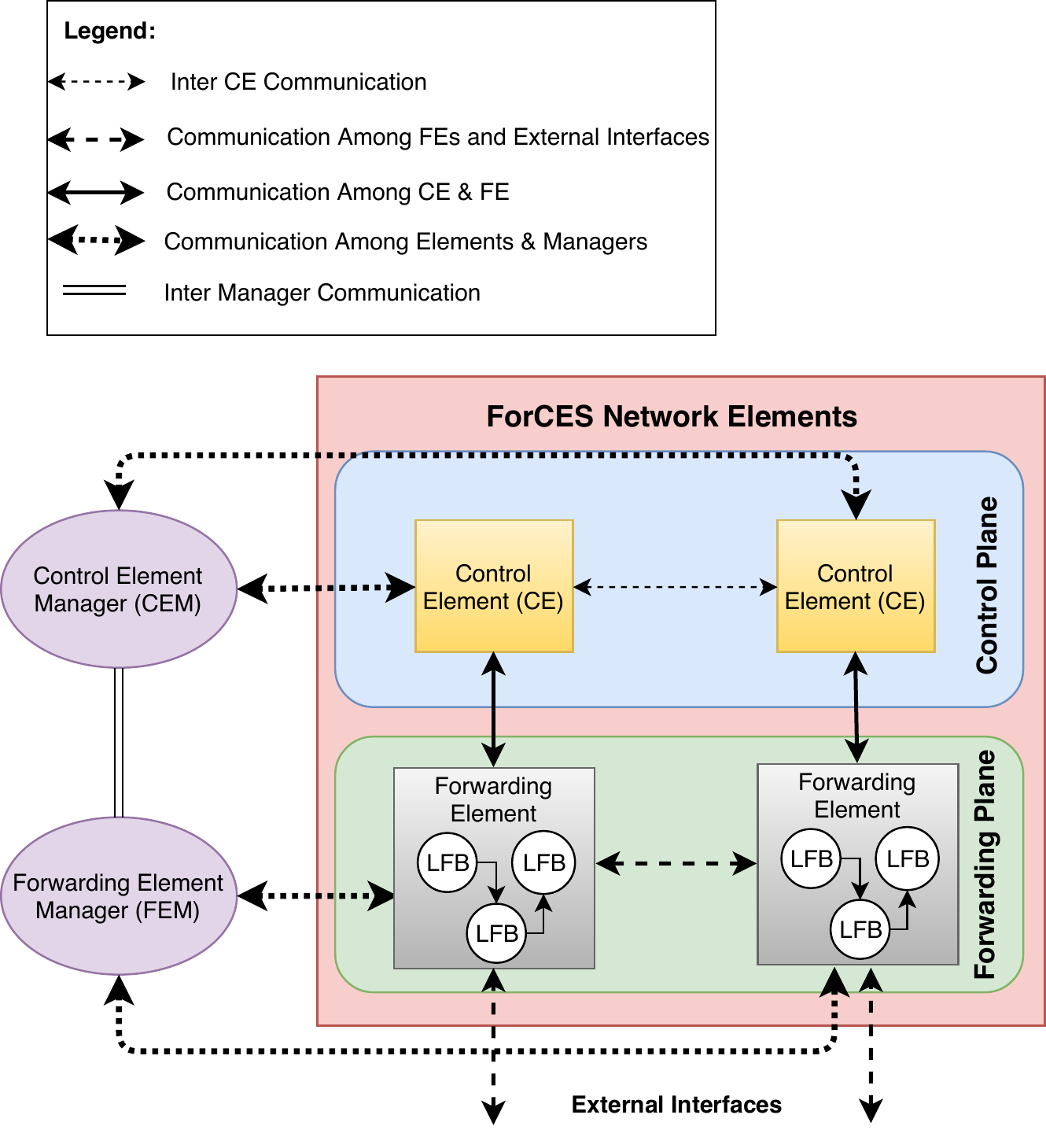}
  \caption{ForCES~\cite{ForCES_Iqbal1} architecture.}
  \label{fig:7}
\end{figure}

\begin{table*}[!t]
	\centering
	\caption{Summary of Comparison Between OpenFlow and ForCES}
	\setlength\tabcolsep{3pt}
	\begin{tabularx}{\linewidth}{|>{\RaggedRight\hsize=1\hsize}X|>{\centering\hsize=1\hsize}X|>{\centering\hsize=0.7\hsize}X|>{\RaggedRight\hsize=1.3\hsize}X|>{\centering\hsize=1\hsize}X|>{\RaggedRight\hsize=1\hsize}X|}
		\hline
		\textbf{Southbound Interface} &
		\textbf{Standardizing Body} &
		\textbf{Protocol Used} &
		\textbf{Determinants} &
		\textbf{Extensibility} &
		\textbf{IP Versions Support}	\\ \hline
		
		ForCES~\cite{ForCES_Iqbal1} &
		IETF~\cite{IETF} &
		SCTP &
		Logical Functional Block &
		Yes &
		IPv4	\\ \hline
		
		OpenFlow~\cite{OpenFlow} &
		ONF~\cite{ONF} &
		TCP &
		Match Fields and Actions &
		No &
		IPv4 and IPv6	\\ \hline
	\end{tabularx}
	\label{tab:3}%
\end{table*}

\begin{table*}[!t]
	\centering
	\caption{Summary of OpenFlow Independent SBI Proposals}
	\setlength\tabcolsep{5pt}
	\begin{tabularx}{\linewidth}{|>{\RaggedRight\hsize=0.4\hsize}X|>{\RaggedRight\hsize=1.4\hsize}X|>{\RaggedRight\hsize=0.8\hsize}X|>{\RaggedRight\hsize=1.4\hsize}X|}
		\hline
		\textbf{Literature} & \textbf{Objective} & \textbf{Solution} & \textbf{Benefits}	\\ \hline
		
		ForCES~\cite{ForCES_Iqbal1} & Separation of Control and Forwarding Element in same Network Element & Logical Function Block & Enables Data and Control Plane separation using traditional network elements  \\ \hline
		
		OpFlex~\cite{OpFlex} & Distribute Complexity and Improved Scalability & Declarative Policy Model & Enhanced Scalability \\ \hline
		
		NetConf~\cite{netconf} & Reduced Complexity and Enhance Performance & Remote Procedure Call (RPC) & Close to native functionality of switch and reduces cost \\ \hline
	\end{tabularx}
	\label{tab:4}%
\end{table*}

Forwarding and Control Element Separation (ForCES) \cite{ForCES_Iqbal1} standardized by IETF \cite{IETF}, is a proposal which is designed to replace OpenFlow. It defines two entities as: Control Element (CE) and Forwarding Element (FE), that are logically kept in same physical device without changing the architecture of traditional networks and without involvement of an external controller as shown in Fig. \ref{fig:7}. It uses a Logical Function Block (LFB) which resides inside FE and has a specific function to process packets and allows CE to control FE. FEs take LFBs as a graph, and uses it to perform well-defined actions and do logical computations on packets which are passing through them. Each LFB can perform a single action on a packet.

ForCES messages are the key enabler to provide the control of FEs to CEs, and just like OpenFlow it also requires a transport protocol. This transport protocol not only provides the communication between FE and CE but also provides some extra services like reliability and security mechanisms. Rather than using TCP for this purpose (as used in OpenFlow), ForCES uses Stream Controlled Transmission Protocol (SCTP) \cite{SCTP} which provides a range of reliability levels. Another major reason for using SCTP is duplication and re-transmission nature of TCP, which in case of congestion, will make things worse. SCTP is also a good design choice for ForCES for it resiliency to failure detection with built in recovery mechanisms.

A detailed comparison between OpenFlow and ForCES is discussed in \cite{OFvsForCES} but in this paper we provide a summarized and condensed comparison between these two competitors. Table \ref{tab:3} summarizes some of the major differences in OpenFlow and ForCES. One of the major difference between OpenFlow and ForCES is that every time a new functionality is added, OpenFlow has to be modeled and standardized accordingly. Whereas, ForCES provides extensibility without need of standardizing again and again. ForCES deployment is not restricted to any specific design of forwarding elements, but for OpenFlow there are switch specifications of predefined features. However, despite being a mature solution, ForCES could not gain widespread adoption by vendors.

\begin{figure}[!t]
	\centering
	\includegraphics[width=0.9\linewidth]{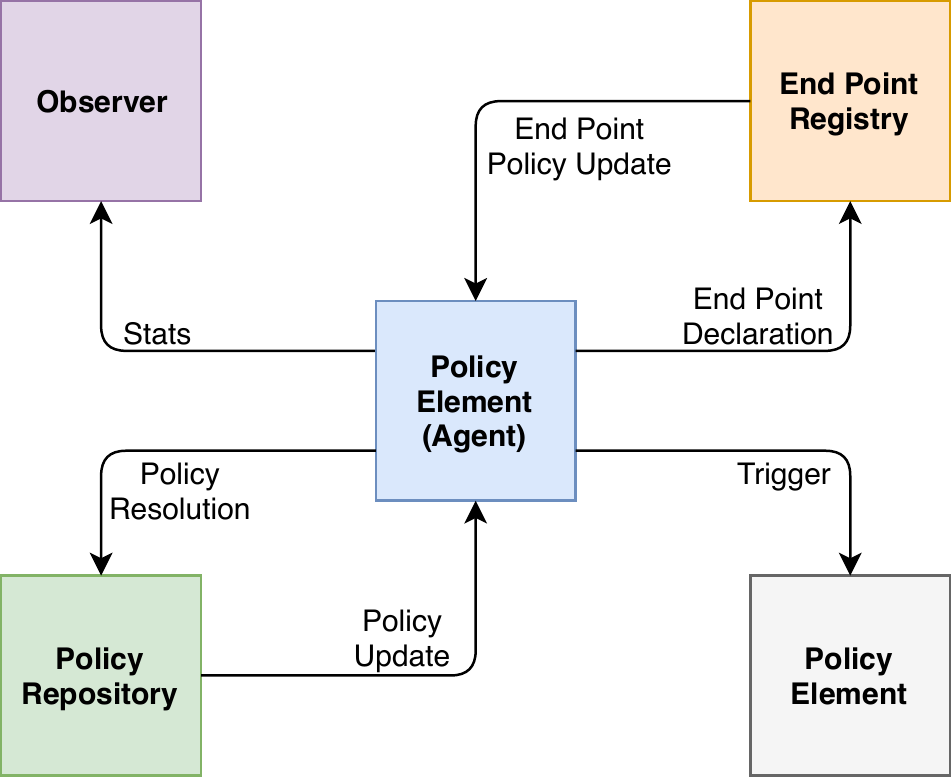}
	\caption{OpFlex~\cite{OpFlex} architecture.}
	\label{fig:8}
\end{figure}

OpFlex \cite{OpFlex}, proposed through a draft for IETF from Cisco, is a protocol that provides communication between centralized controller and data plane but with a very different scope as compared to OpenFlow. OpFlex is based on declarative policy information model that means it centralizes only policy management and implementation, but distributes intelligence and control. With the aim of scalability, OpFlex tries to distribute the complexity in such a way that forwarding devices are responsible for managing the whole network except policies. These policies are defined at logically centralized Policy Repository (PR) which communicates with Policy Elements (PE) using OpFlex protocol. End Point devices are connected with Policy Elements and get registered by using End Point Registry which is responsible for the addition/removal of End Points. Another repository in OpFlex is Observer (OB) which is responsible for statistics faults and events. This interaction is depicted in Fig. \ref{fig:8}. One major limitation of OpFlex, as compared to OpenFlow, is that it takes away the key feature of programming the network from a centralized controller.

NetConf~\cite{netconf} which uses Remote Procedure Call (RPC) paradigm is a protocol that defines a simple mechanism by which network devices can be managed, configuration data can be retrieved, and new configuration data can be uploaded and manipulated. One of the key aspect of NetConf is that it closely mirrors the functionality of the management protocol to the native functionality of the device which directly reduces cost and allows timely access to new features. This proposal existed before SDN, but just like OpenFlow it also provides a straightforward API. This API can be used by applications to send and receive full or partial configuration datasets.

\textbf{Conclusion:}
Some major benefits of ForCES over OpenFlow are its extensibility as well as no restriction on device specification. It offers a rich set of features but lacks in open source support. Similarly, OpFlex restricts programming in networks which is a key feature of SDN. NetConf, on the other hand, is not a purpose built interface for SDN, thus it does not provide enough flexibility. Table \ref{tab:4} presents a summary of all the OpenFlow independent proposals with their objectives, solutions, and benefits. It is interesting to note that although ForCES was supported by IETF, but it did not gain the industry confidence. The major reason was vast deployment of OF and support from multiple developers. 

\subsection{Southbound APIs for Wireless Sensor Networks} 
Application of SDN has expanded to different types of networks, one of which is Wireless Sensor Networks \cite{SDWSN}. In this section we discuss SDN and use of APIs in WSNs. Development of smart sensors has enabled the deployment of sensor networks in the past decade. They are capable of monitoring physical and environmental factors. Software Defined Wireless Sensor Networking (SDWSN) is a new paradigm for Low Rate-Wireless Personal Area Network (LR-WPAN) which can be realized by infusing SDN model into WSN. Southbound Interface (e.g. OpenFlow) plays an integral role in SDN, but it is very hard to implement in SDWSN because of the following reasons:

\begin{itemize}
\item Matching fields in OpenFlow are address centric, and flow entries are installed using sourceIP, destinationIP. Whereas, WSN are data centric, where data acquisition is more important than source of data. Hence, flow creation is challenging in WSNs.
\item Addressing in WSN is not IP-based which prevents SDWSN SBI from creating flow entries. Moreover, it becomes hard to establish a TCP/IP based secure channel in SDWSN.
\item To install or uninstall flows on sensor devices which are limited in size and memory, it may introduce overhead on communication channel.
\item Due to the device constrains, routing algorithms of WSN are quite different than data centers or other networks. Hence, topological information needed is more detailed and may not be represented by OpenFlow headers. 
\end{itemize}

Sensor OpenFlow (SOF) \cite{SOF} is based on standard OpenFlow, but modified to the requirements of low capacity sensor nodes. It addresses challenges like; flow creation, secure channel between control plane and data plane, control traffic overhead, and in-network processing, etc. To install the flows on sensor network devices, it suggests to redefine the flow tables due to special addressing schemes of WSN. Flow tables are categorized in two classes: Class1 contains compact network unique addresses as 16 bit addresses in ZigBee, and Class2 uses concatenated attribute value pairs. Class1 is handled by use of OpenFlow eXtensible Match (OXM), a TLV format by adding two new addresses as OXM\_SOF\_Source and OXM\_SOF\_Destination. Another solution for this problem is to use \textit{uIP} and \textit{uIPv6}. \textit{uIP} is an implementation of IPv4 in Contiki operating system  normally used for WSN and Internet of Things. Just like an OpenFlow secure channel, SOF suggests either to use Transport Protocol directly to WSN or channels can be supplied through \textit{uIP or uIPv6}. To curb the control traffic between data and control plane, it proposed a customized solution named as Control-Message Quenching (CMQ). However, the main focus was on message type, packet format, and operations, and did not provide any performance evaluation.

Software Defined Wireless Networks (SDWNs) \cite{SDWN} proposed some significant features to reduce energy consumption in WSN by introducing duty cycles and in-network data aggregation. Another significant feature of SDWN is to support flexible definition of rules. Duty cycles are used to reduce the energy consumption by turning radio off when it is not being used. Another approach used to reduce energy consumption is in-network data aggregation. Unlike traditional OpenFlow, flexible flow entries are required for SDWN because of its nature. SDWN protocol architecture uses generic nodes as well as sink node. All generic nodes run physical and MAC layer functionalities. Forwarding layer which is on top of MAC layer is responsible to treat a packet as specified by controller. All the generic nodes are connected to sink node(s) which has same architecture as generic nodes except a few functionalities. A sink node has more computational and communication capabilities. Therefore, sinks are executed in Linux based embedded system. Embedded system and sinks are connected through USB, RS232, or other interfaces. Another feature of sink is to use virtualizer with the responsibility of collecting information of generic nodes to build a detailed representation of network topology. Same as SOF, SDWN also did not provide any performance evaluation and mainly focused on architectural details. Hence, actual performance is still unknown and may be a research direction for community.

SDN for WIreless SEnsors (SDN WISE) \cite{SDNWISE} goes one step ahead as compared to previous studies, and is implemented in OMNet++ with the objectives of reducing communication among sensor nodes to/from SDN controller and making sensor nodes programmable as finite state machine unlike standard OpenFlow which is stateless. In control layer, it uses WISE Visor, which has Topology Manager (TM) for collection of local information from the nodes and forwards it to the controller in the form of graph with the topographic information, energy levels, and SNR of nodes. In data plane In-Networking Packet Processing (INPP) is responsible for data aggregation and other in-network processing to reduce the overhead. Between control and data plane there is adaptation layer which is responsible for formatting the messages received from sinks in such a way that they can be handled by WISE-Visor and vice versa. An application of SDN WISE is in \cite{BeyondvSwitch} where a unified system is realized, which enables communication of heterogeneous devices under a single Network Operating System by adding subsystems like Sensor Node, Sensor Flow Rules and Sensor Packet Subsystems.
\begin{table}[!t]
	\centering
	\caption{Summary of SBI Proposals for Wireless Sensor Networks}
	\setlength\tabcolsep{3pt}
	\begin{tabularx}{\linewidth}{|>{\RaggedRight\hsize=1.5\hsize}X|>{\centering\hsize=0.7\hsize}X|>{\centering\hsize=0.9\hsize}X|>{\hsize=0.9\hsize}Z|}
		\hline
		\textbf{Features}	& \textbf{SOF~\cite{SOF}}	& \textbf{SDWN~\cite{SDWN}}	& \textbf{SDN-WISE \cite{SDNWISE}}	\\ \hline	    
		\textbf{Flow Creation}	&Yes	&Yes	&Yes	\\ \hline
		\textbf{Field Matching} &Yes	&Yes	&Yes	\\ \hline
		\textbf{Action}			&No		&Yes	&Yes	\\ \hline
		\textbf{Statistics}		&No		&No		&Yes	\\ \hline
		\textbf{Data Aggregation}	&No	&Yes	&Yes	\\ \hline
		\textbf{In-Network Processing}	&Yes	&Yes	&Yes	\\ \hline
		\textbf{Duty Cycles Reductions}	&No		&Yes	&Yes	\\ \hline
		\textbf{Mobility Management}	&No		&No		&No		\\ \hline
		\textbf{Implementation Available}	&No	&No		&Yes	\\ \hline
	\end{tabularx}
	\label{tab:5}%
\end{table}

{\bfseries{Conclusion:}} Table \ref{tab:5} summarizes the feature based comparison of different southbound interfaces for WSNs proposals. SOF and SDWN provide theoretical details using a centralized controller, whereas SDN-WISE provided practical implementations using ONOS controller which is distributed. Nodes in sensor networks are susceptible to movement which can cause path variation during packet transmission. It is very important to manage and monitor the movement of different nodes. One of the major challenge of SDN is to handle the effect of nodes entering or leaving the network. another challenge is to build paths using different metrics (i.e. node energy and capability). The interface in such cases should be able to optimally gather required information for the controller. 

\subsection{Southbound APIs for Internet of Things} 
Smart cities, smart grids, and intelligent transportation has expanded the Internet of Things domain significantly. IoT networks are not just WSN, in fact they are more complex and implement WSN as a sub-part of the whole ecosystem. Due to a large number of devices connected to Internet, there are a number of challenges in IoT: scalability, connectivity, big data, security and heterogeneity, etc. SDN provides a centralized controller and high level management which  hides the complexity to provide solutions for above discussed problems. Implementation of SDN in IoT networks resolves a number of issues as well introduces some new challenges. 
\begin{itemize}
	\item {\textbf{Device Heterogeneity:}} IoT devices are very diverse in nature, and may use different types of technologies. Their capabilities also may vary, which requires new types of software defined solutions, including controllers, vswitch/SDGateways, and southbound interfaces.

	\item \textbf{Interface and Topological Diversity:} Each IoT device may have multiple communication technologies, e.g. WiFi, BLE, 5G, etc. As the flow installation on such network is not simple, hence the southbound interfaces has to adapt. Moreover,  SBIs also should be able to work with hybrid wired and multi-hop wireless networks.

	\item {\textbf{Protocol Integration:}} Each technology in IoT may have its own packet format and processing rules. Flow installation with such a variety of protocols is a very challenging task.
\end{itemize}

\begin{table*}[!t]
	\centering
	\caption{Summary of proposals for Internet of Things}
	\setlength\tabcolsep{3pt}
	\begin{tabularx}{\linewidth}{|>{\RaggedRight\hsize=0.7\hsize}X|>{\RaggedRight\hsize=1.1\hsize}X|>{\RaggedRight\hsize=1.1\hsize}X|>{\RaggedRight\hsize=1.1\hsize}X|>{\RaggedRight\hsize=0.8\hsize}X|>{\RaggedRight\hsize=1\hsize}X|>{\RaggedRight\hsize=1.2\hsize}X|}
		\hline
	    \textbf{Literature} &
	    \textbf{Objective} &
	    \textbf{Solution} &
	    \textbf{Benefits} &
	    \textbf{SBI Used} &
		\textbf{Changes in Southbound Interface} &
		\textbf{Limitations}
	    \\ \hline
	    
	    Salman~et~al. \cite{Salman} &
	      Issues of scalability, reliability and heterogeneity &
	      Implementation of gateways running genius algorithm &
	      Provides control on IoT devices by extending OpenFlow &
	      OpenFlow &
	      Not discussed &
	      No implementation available. Extensions in OpenFlow are not discussed properly
		\\ \hline
	    Li~et~al. \cite{Yuhong} &
	      Resource sharing and interoperability &
	      Data processing and storage center &
	      Supports multiple services and interoperability &
	      OpenFlow &
	      Addition of two fields in OpenFlow header &
	      Security designs are not focused
		\\ \hline
	    Ojo~et`al. \cite{Ojo} &
	      Scalability and Mobility &
	      Use of Software Defined Gateways instead of traditional Gateways &
	      Enhances network efficiency and agility &
	      OpenFlow, OvSDB, BGP, PCEP, NetConf &
	      Not discussed &
	      No implementations available. Extensions in southbound protocols are not discussed
		\\ \hline
	    Qin~et~al. \cite{MINA} &
	      Heterogeneous nature of different technologies and Interoperability &
	      IoT Multi-network Controller &
	      Provides flexible, effective and efficient management &
	      OpenFlow like Protocol &
	      Not discussed &
	      Lacks in details of southbound protocol
		\\ \hline
	    Desai~et~al. \cite{IoTOF} &
	      Better Control on IoT Heterogeneous Devices &
	      Introduced OpenFlow enabled management device using Linux kernal &
	      Supports heterogeneous IoT devices to communicate with Remote Processing systems in cloud &
	      OpenFlow &
	      Not discussed &
	      No discussion about flow installation
	  \\ \hline
  \end{tabularx}
  \label{tab:6}%
\end{table*}

To address these challenges, an OpenFlow like solution is required for IoT using SDN. There are a number of solutions available in literature, but most of them do not resolve all the problems. Table \ref{tab:6} presents a summery of these proposals.

Salman et al. \cite{Salman} proposed an architecture for implementing SDN in IoT and proposed a layered architecture to overcome problems like; scalability, big data, heterogeneity and security. Bottom most layer is device layer with different IoT devices and identifiers to differentiate them. Network layer is used to overcome the heterogeneity by using Software Defined Gateways (SD Gateways). SD Gateways can communicate with IoT devices using different technologies. An extension to OpenFlow is recommended but no specifications are discussed. However, for configuration purposes a number of management protocols (e.g. NetConf, OF-Config, and Yang) are recommended. Another main feature of these SD Gateways is to reduce the power consumption because of big data problem. Control layer consist of SDN controllers with the responsibility of collecting topology information, path calculation, and forwarding rules. Security rules are defined using algorithms but no details on flow rule installation is discussed. At the top most layer, there are different network applications.

Li et al. \cite{Yuhong} address the issues of interoperability, resource sharing, and flexibility for applications and services. It proposes a layered architecture where IoT devices are at bottom layer and is referred as device layer. These devices are connected to switches or gateways which are at communication layer. A module named as Data Processing and Storage Center is also in communication layer and controlled by SDN controller. This module is capable of storing selective data of IoT devices and sinks, and also responsible for data format conversion. On top of communication layer, there is computing layer where SDN controllers are placed. Service layer is the top most layer. Besides data forwarding capability of switches and gateways, they can also store or cache local data and process it under the instruction of SDN controller. An extension in OpenFlow version 1.0 is done by adding two flags. These flags mark data format and caching capabilities of the switch.

To resolve the problems of scalability and mobility, Ojo et al. \cite{Ojo} provide a general architecture for IoT with the coupling of SDN and NFV. This architecture contains four layers; perception layer, data layer, control layer and application layer. Devices in perception layer sense data and forward to data layer by using Software Defined enabled gateways. These Software Defined enabled gateways provide management flexibility, as underlying devices belong to different technologies. Apart from these Software Defined gateways, there are also switches in data plane. These devices can be programmed by through controllers (e.g. ONOS, ODL) by using southbound interface (e.g. OpenFlow, OvSDB, NetConf, BGP etc.). Application layer is on top of control layer, from where different services can be implemented. This study lacks implementations and description about flow installations and the device are controlled by given southbound interfaces.

Multi network INformation Architecture (MINA) \cite{MINA} is another method to resolve the issues of heterogeneity and interoperability. It proposes a controller architecture and an OpenFlow like protocol. In the controller, it uses data collection components which collect network information and stores it in databases. This information is then utilized by other components of controller. Among these components, there is an admin/analyst API which allows to govern different control processes by controller itself as well as external programs. Other components are; task-resource matching, service solution specification, and flow scheduling. A task can be realized by single service or multiple services. Task-resource matching specifies, which devices or applications can be used to complete a particular task. After matching, controller maps the characteristics of devices and services involved in that matching by using service solution specification component. It also handles specific requirements for devices or application constraints. These requirements are taken by the flow scheduling component to schedule flows. This component uses an algorithm to resolve the complexity due to heterogeneity of different technologies. An OpenFlow like protocol is used in communication layer for flow scheduling and data collection purposes. However, detailed discussion and working of it is not discussed in the paper.

In \cite{IoTOF} Desai et al. provided a framework where an OpenFlow Management Device is responsible to provide communication between IoT devices and OpenFlow enabled switches. This device runs its own Linux based operating system. Bottom layer consists of hardware and protocols and these components are the base of this device. It is an extensible device and allows to add more protocols. Above this layer, there are libraries which provide different functionalities like; security, web connection, and SQL. Application framework layer is on top of libraries with resource manager, location manager, activity manager, and above this there is an application layer. Data plane devices communicate with control plane as well as with this device using OpenFlow protocol. Flow installation mechanisms, in this study, are not discussed.

{\bfseries{Conclusion:}}
Table \ref{tab:6} gives a comparative analysis of the proposals discussed in this section. Most of the literature related to IoT has to focus on two parts: controller and API. We find that more focus is given to controller design and it is assumed that OpenFlow or something similar will be able to communicate with the devices. The objective of this paper is limited to APIs, hence we limit this section to these works which have elaborated (even in passing) on the SBIs. It is important to highlight the necessity SBIs specific for IoT devices. OF was not designed for mobile low capacity heterogeneous IoT devices. Hence, firstly it is important to evaluate the effect of OF communication performance in such networks, and then perhaps a more lightweight and customized API can be developed targeted for IoT networks. In IoT, OpenFlow is not limited to controller and vSwitch. It has to extend its reach to IoT devices. Hence, solutions which go beyond the SDN gateways is an important research area. Similarly, the SBI also needs to offer functionality other than flow installation.

\section{Northbound Interfaces (NBI) in SDN}

Northbound Interface is one of the key pillars of SDN, as it provides programming abstraction for networks. It acts as a bridge between control and management plane, and provides high level abstraction for application development. The application development in management plane is not as easy as it should be, and the main reason for it is the lack of standardization of northbound interface. Unlike OpenFlow, there is no single API or protocol which different developers/vendors can use. One reason for lack of this standardization is the variation in applications and their requirements. Northbound Interface Work Group (NBIWG) \cite{NBIWG} is an initiative of ONF \cite{ONF}, which has been established for standardization purposes. Because of the absence of a standardized NBI, there are some controllers (e.g. Onix \cite{Onix}, PANE \cite{PANE} etc) which provide a higher level API for their application development in SDN.
Furthermore, programmers and most of the controllers usually use REST API as Northbound Interface.


\begin{table*}[!t]
  \centering
  \caption{Summary of Portability solutions in NBI Proposals}
	\setlength\tabcolsep{3pt}
	\begin{tabularx}{\linewidth}{|>{\RaggedRight\hsize=0.6\hsize}X|>{\RaggedRight\hsize=1.1\hsize}X|>{\RaggedRight\hsize=1.1\hsize}X|>{\RaggedRight\hsize=1.2\hsize}X|}
		\hline
    	\textbf{Literature}	& \textbf{Objective}	&\textbf{Solution}	&\textbf{Benefits}	\\ \hline
    	
    	SFNet~\cite{SFNet}	& Interaction among applications and underlying devices & 
    		Uses plug-ins for various applications	&
    		Allows bandwidth reservations, multi-casting and supports congestion inquiry  \\ \hline
     	
     	tinyNBI~\cite{tinyNBI}	& Foundational interface for OpenFlow versions	&
     		Uses data model to abstract specifications	&
     		Supports multiple version of OpenFlow and provides extensibility  \\ \hline
     		
    	NOSIX~\cite{NOSIX}	& Switch diversity and performance enhancement	&
    		Virtual Flow Table (VFT) and Switch Driver	&
    		Provides flexibility to programmers for a diverse landscape of data plane devices  \\ \hline
	\end{tabularx}
	\label{tab:7}%
\end{table*}


The working of SBIs are more like a protocol for communication, whereas NBIs are used for different objectives. \cite{7889501} identifies some key properties of NBIs. Using it as a rudimentary, we group the literature of northbound interfaces on the following properties: portability, programmability, controller based, intent based, and virtualization.

Portability provides low level abstraction and is used to resolve the compatibility issues among different versions of OpenFlow or other southbound APIs and different hardware. It provides the guarantee of correct packet processing on wide range of data plane devices. 
Programmability refers to the use of high level programming languages or dedicated languages for SDN. By using high level languages, networks can be configured for the services required, which is referred as prescribed usage. Whereas, there is an intent based usage of NBIs, which in totally opposite to prescribed usage. In this model, application requirements are described in natural language and controller is intelligent enough to integrate desired services with its core functionality.

Due to the absence of a standardized northbound interface, many of the controllers use ad-hoc APIs. Whereas, some of the controllers proposed their own high level interfaces referred as controller based APIs. Moreover, some of the controllers also support intents where high level policies can be declared.
Interfaces allow the sharing of resources of underlying network devices in virtualization and reduces capital and operational costs. Due to the blurriness of virtualization techniques in SDN we have discussed virtualization as a separate section. In this section we discuss portability, programmability, controller based, and intent based NBIs. 


\subsection{Portability in NBI}
A reason which hinders application development is a gap between hardware vendors and application developers which reduces portability i.e. guarantee of correct packet processing and good performance over a wide range of network switches. There are a large number of different (hardware and software) switch products available from dozens of vendors, which differ in data plane, switch-controller interaction, and fixed/flexible pipeline \cite{vendors}.

Software Friendly Networking (SFNet) \cite{SFNet} provides an interface between control and application plane of SDN paradigm with a role to hide the lower network protocols from the application. It is a high level API and directly interacts with the underlying network. It translates application requirements and programs network accordingly to provide services. It uses JSON file to send information requests and congestion report from the network to find a better path among hosts. In case of congestion, it allows applications to back off if they choose to do so. It can be beneficial in case of delay sensitive traffic. It also supports bandwidth reservation and grants/denies the request depending upon the availability of requested bandwidth. This requirement of bandwidth can be prioritized for video on demand or Voice over IP (VoIP) traffic. It also supports multicasting to a set of IP addresses participating in it.

NOSIX \cite{NOSIX} addresses the problem of diverse range of underlying switches to enhance the performance. It proposes use of Virtual Flow Table (VFT) which is a basic component used by applications to freely define the rules without any concern of delays and throughput of updates and notifications. It allows applications to predefine the VFTs pipeline and then install rules in these tables. Predefinition of pipeline is allowed because it is very hard to do dynamic reconfiguration of physical pipeline of switches. Moreover, the rules in VFTs do not need to be always in physical flow tables of a switch. Switch drivers, on the other hand, maps the VFT pipeline onto a physical pipeline available on the switch. Switch driver can be placed either at the lower layer of controller or on the switch itself. However, it is beneficial to place switch drivers at lower stack of controller because it is easy to program software based switch driver at the controller than at switch. In NOSIX, control applications can be written as a pipeline of VFTs and vendor supplied drivers then transform them into switch configuration.

Another solution is tinyNBI \cite{tinyNBI} which is language independent solution and resolves portability issues in terms of different OpenFlow versions and specifications. It is designed in C language, and provides a complete set of OpenFlow semantics and handles multiple versions of OpenFlow and variable switch capabilities without requiring any additional efforts.  Main purpose of this NBI is to provide a foundational interface for application development. It uses a data model which makes a clear distinction between control and data plane abstractions. Every abstraction has three components; configuration, capabilities and statistics. Configuration is modifiable data by the interface. Capabilities describes the behavior of abstractions and it is non-modifiable state. Whereas, statistics are read-only data and describes how abstraction has behaved. All the abstractions are not present in all OpenFlow version (e.g. Meter Table, Group Table). Abstractions which are not available are handled in three ways; seamless emulation, switch offloading, and error indication. Seamless emulation defines the abstraction but with limited capabilities. Switch offloading represents offloading of missing switch capabilities to controller. In some scenarios, it is not possible to provide missing behavior which provides an error indication. It is not a high level interface and does not provide information like network topology, switching, routing, or load balancing. Instead, it provides independence from ever changing structure and semantics of OpenFlow and provides maximum portability and re-use potential.

{\bfseries{Conclusion: }}
One of the major responsibility of NBI is to provide the information of underlying devices to developers. However, there is a diverse range of underlying devices and southbound protocols. Portability provides solutions for compatibility issues of this diversity of network elements and protocols. Table \ref{tab:7} presents a summary of different proposals for portability in NBIs. These solutions focus on either data plane devices or protocols, but a single solution for both of these is still an open research challenge.

\begin{figure*}[!t]
	\centering
	\includegraphics[width=0.95\linewidth]{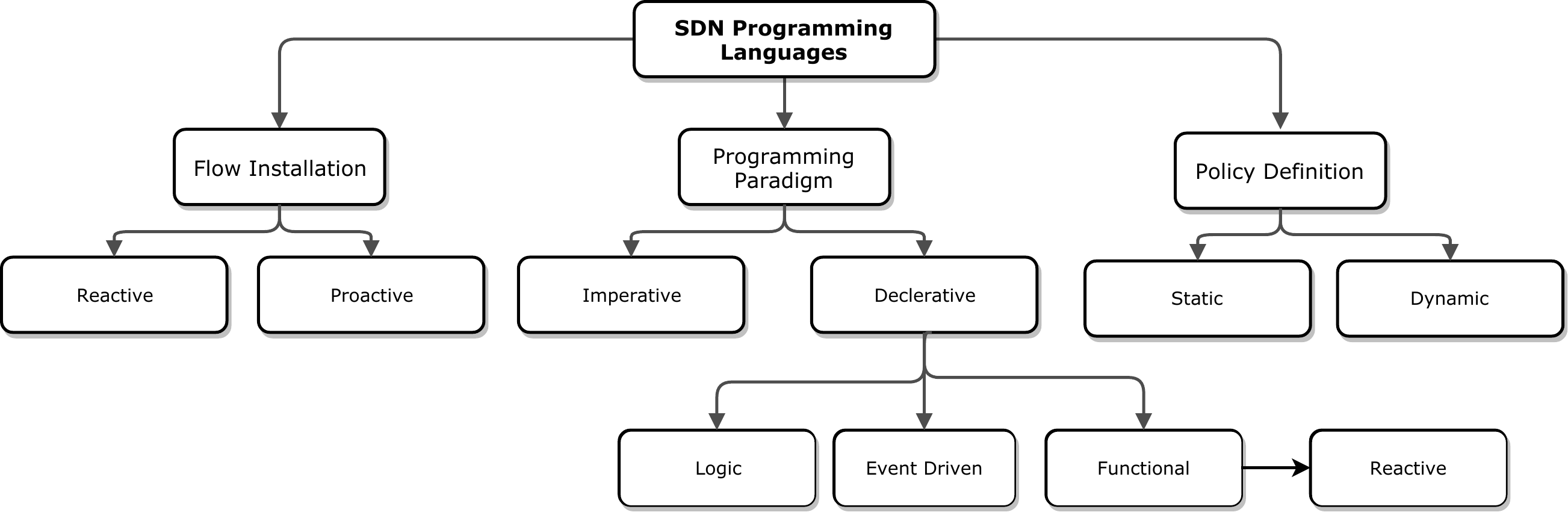}
	\caption{Feature classification of NBI programming languages in SDN.}
	\label{fig:9}  
\end{figure*}

\subsection{Programmability of NBI}
Similar to application development, which shifted from assembly language to high level languages like Python and Java, network languages have also shifted from low level language (e.g. low level language used in OpenFlow) to high level (e.g. Frenetic, Procera etc.). These high level languages in SDN provides flexibility in network management and make it less error prone.
 
Current network elements often perform multiple tasks simultaneously (e.g. routing, monitoring, access control, etc.). However, decoupling these tasks is almost impossible, because packet handling rules installed by one can be in conflict with another. OpenFlow interface is defined at very low level of abstraction which directs capabilities of switch hardware. To apply high level concepts, programmers can not directly use OpenFlow instruction set. Another issue with OpenFlow is that, packets are processed by controller if switch can not process them due to lack of flow information, hence programmers have to perform two tier programming, one for the packets being processed at controller and other which need to be processed at switch level. The literature related to NBI programming languages has been discussed in following sub-section. Prior to it, we have identified different properties of such languages.

\subsubsection{\bfseries{NBI Programming Language Feature Classification}}
The feature classification of NBI-PL is shown in Fig. \ref{fig:9}. Here we give brief description of each feature.

\textit{\bfseries{Flow Installation}} refers to the way in which forwarding rules can be installed on switches: i.e. \textit{Reactive} and \textit{Proactive}. Almost all of languages provide reactive approach, however, some additionally provide proactive flow installation capabilities. In reactive approach, when a new packet arrives at switch, and no flow information is available, then this packet is forwarded to the controller. It based on its program logic installs flows in the flow table of switch. This method introduces latency as every new packet will be sent to controller for flow installation. On the other hand, proactive approach eliminates latency as every packet is not sent to controller for forwarding rules. In fact, predefined rules and actions are installed in flow tables. Languages that perform proactive installation, pre-compute forwarding tables for the whole network, and controller only modifies flow tables in case of link failure or other external events.

\textit{\bfseries{Policy Definition}} is a set of aggregated rules for the network, and each policy is a set of conditions and a corresponding set of actions where condition defines that which policy rule will be applied. Policy definitions  are classified to \textit{static} and \textit{dynamic} policies. Static policies (most traditional methods for firewall filters) are predefined set of rules and actions. Dynamic policies, on the other hand, can be changed according to network conditions. Although most languages have dynamic capabilities, but there are some which only work with static policy definition.

\textit{\bfseries{Programming Paradigm}} reflects different ways of building the structure and elements of any program. There are two different paradigms in SDN programming: \textit{imperative} and \textit{declarative}. Imperative paradigm can be viewed as traditional programming structure and allows the programmer to specify all the steps to solve any particular problem. In declarative paradigm, one just need to specify what program must do, not how to do it \cite{declarative}. The most known example of declarative paradigm is Structured Query Language (SQL), where a query is stated and database engine executes it. Declarative paradigm has further three sub-paradigms: logic programming, event driven, and functional programming. In \textit{Logic Programming}, compiler applies an algorithm that scans all possible combinations to a set of defined inference rules to postulates and resolves a query. \textit{Event Driven Programming} allows the program to respond to any particular event. As soon as an event is received, an automatic action is triggered. This action can either be some computation or trigger another event. \textit{Functional Programming} acts as an evaluation of some mathematical functions and avoid state changes. The\textit{ Reactive Programming} in declarative paradigm facilitates programs to react on external events. For example, a spreadsheet which typically has values or formulas in its cells. Whenever cell changes, formulas are recalculated automatically. A combination of functional and reactive programming makes \textit{Functional Reactive Programming} (FRP) which models reactive behavior in functional languages. Programs in FRP correspond to mathematical functions in a declarative manner.

\begin{table*}[!t]
	\centering
	\caption{Feature Based Summery of SDN Languages for NBIs}
	\setlength\tabcolsep{3pt}
	\begin{tabularx}{\linewidth}{|>{\RaggedRight\hsize=0.6\hsize}Y|>{\centering\hsize=0.7\hsize}X|>{\centering\hsize=0.7\hsize}X|>{\centering\hsize=0.8\hsize}X|>{\centering\hsize=0.8\hsize}X|>{\RaggedRight\hsize=2.4\hsize}X|}
		\hline
		
		&
		\multicolumn{2}{>{\centering\setlength{\hsize}{1.4\hsize}\addtolength{\hsize}{2.1\tabcolsep}}X|}{\textbf{Flow Installation}} &
		&
		\textbf{Programming}	&	\\\cline{2-3} 
		
		\multirow{-2}{*}{\textbf{Literature}}	&
		\textbf{Reactive}	&\textbf{Proactive}	&
		\multirow{-2}{*}{\textbf{Policy Definition}}	&
		\textbf{Paradigm}	&
		\multirow{-2}{*}{\textbf{Description}}	\\ \hline
		
		Frenetic~\cite{Frenetic}	& {\checkmark}	& -	& D	& FR	& 
		Designed to avoid race condition by using well defined high level programming abstraction	\\ \hline
		
		FML~\cite{FML}		& {\checkmark}	& -	& S	& L		& 
		Designed for policy definition in enterprise networks		\\ \hline
		
		Flog~\cite{Flog}	& {\checkmark}	& -	& D	& L/ED	&
		Executes programs on event occurrence in network			\\ \hline
		
		Procera~\cite{Procera}	& {\checkmark}	& -	& D	& FR	&
		An expressive language and used for policy handling and provide an extensible and compositional framework	\\ \hline
		
		Pyretic~\cite{Pyretic}	& {\checkmark}	& {\checkmark}	& D	& I	& 
		An upgrade of Frentic and also used for policy handling in a transparent framework		\\ \hline
		
		Nettle~\cite{Nettle}	& {\checkmark}	& {\checkmark}	& D	& FR	&
		Allows programmers to deal with streams instead of events and can be understood as signal functions			\\ \hline
		
		NetKAT~\cite{NetKAT}	& {\checkmark}	& {\checkmark}	& D	& F	&
		Provides reasoning for network mapping and traffic isolation using Kleene Algebra and Tests					\\ \hline
		
		NetCore~\cite{NetCore}	& {\checkmark}	& {\checkmark}	& D	& FR	&
		Provides means for packet forwarding policies and generate flow installation commands	\\ \hline
		
		FlowLog~\cite{FlowLog}	& {\checkmark}	& {\checkmark}	& D	& F	&
		Allows programmers to use external full featured libraries		\\ \hline
		
		FatTire~\cite{FatTire}	& {\checkmark}	& {\checkmark}	& D	& F	&
		Provides fault tolerance in the networks and describe network paths						\\ \hline
		
		Kinetic~\cite{Kinetic}	& {\checkmark}	& {\checkmark}	& D	& ED	&
		Domain Specific Language that allows to control the network dynamically					\\ \hline
		
		Merlin~\cite{Merlin}	& {\checkmark}	& -	& D	& F	&
		Delegates sub-policies to different tenants and allow them to modify according to their requirements		\\ \hline
		\multicolumn{6}{>{\centering\setlength{\hsize}{6\hsize}\addtolength{\hsize}{10.95\tabcolsep}}X}{Legend: D=Dynamic, S=Static, I=Imperative, ED=Event Driven, FR=Functional Reactive, L=Logic, F=Functional}	\\ 
	\end{tabularx}
	\label{tab:8}%
\end{table*}

\subsubsection{\bfseries{Programming Languages for NBI}}
Here we present a list of programming languages specifically designed for SDN usage. Table \ref{tab:8} presents a listing of classification features of these languages.

Frenetic \cite{Frenetic}, embedded in Python, proposes two levels of abstraction which includes a set of source level operators for constructing and manipulating streams of network traffic, and a declarative system that handles all the details of installing and un-installing low level rules on switches. It provides a declarative solution with modular design. Using Frenetic, programmers do not need to be concerned about flow rule installation which may prevent controller from analyzing other traffic. Flow based Management Language (FML) \cite{FML} is a high level declarative language based on data log for policy configuration about a verity of management tasks within a single framework for large enterprise networks. Main issue with FML is that it applies policy on all packets of any particular flow and does not provide much flexibility. Flog \cite{Flog} is another event driven and forward chaining language for SDN which combines the idea of FML and Frenetic as it follows logic programming technique as FML and factored programs in three components like Frenetic: a method to query network state, a component to process data generated after queries, and a mechanism to generate rules for installing on network elements. It uses event driven and logic paradigms, and programs in this language execute upon occurrence of an event in network.

To design the network policies, a language must be expressive enough to capture these policies. Procera \cite{Procera}, as compared to FML, is more expressive and handles policies in a better manner. It allows network operators to define policies which react to dynamic changes in different network conditions. It provides an extensible, expressive, and compositional framework. An up-gradation of Frenetic is proposed by same developers as Pyretic \cite{Pyretic}, which is a Python based platform that allows application developers to design sophisticated applications. Same as Procera it also helps in policy based application design. It tries to resolve the shortcoming in OpenFlow in terms of its programming nature and its role as a programming interface to switch in the network. Policies in Pyretic support modular programming and also facilitates creation of dynamic policies.

NetCore \cite{NetCore} provides packet forwarding policies in SDN which is expressive, compositional, and has formal semantics. Rather than using the SDN controller, NetCore provides compilation algorithms and couples them with the run time system which issues flow installation rule commands. 
It exclusively focuses on flow tables simplicity but lacks expressiveness. To solve this issue, FlowLog \cite{FlowLog} which is a tier-less programming language, provides flexibility to use external full featured libraries. Nettle \cite{Nettle} using the mantra \textit{"Don't Configure the Network, Program it!"} adopts ideas from Function Reactive Programming (FRP) and design methodology from Domain Specific Language (DSL), is embedded in strongly typed language Haskell (which works as a host language). It can be understood as signal functions used in electrical signals and provides flexibility to change these functions as well as retrieve discrete time and contentious time values.

Another proposal for network programming in SDN is NetKAT \cite{NetKAT} which uses mathematical structure called Kleene Algebra for tests and provides solid mathematical semantic foundations. It provides equational theory for reachability, traffic isolation and compiler correctness of algorithms.

FatTire \cite{FatTire} provides forwarding and fault tolerance policies. It allows programmers to set legal paths through the network along with fault tolerance of those paths. Network conditions are always changing and operators have to change the configuration manually. Kinetic \cite{Kinetic}, based on Pyretic, proposed an intuitive mechanism to change these configurations dynamically. It expresses network policy as a Finite State Machine (FSM) which captures dynamics and amenable to verifications. It also verifies the correctness of these high level specifications.

To manage the networks, administrators need to configure their network very carefully because misconfiguration of a single device may bring an undesired behavior to the whole network. By using Merlin \cite{Merlin}, administrators can express policies in high level declarative language. Merlin compiler uses program partitioning to transform global policies to smaller sub-policies which are distributed to different components of network automatically, and delegates these sub-policies to different tenants who can modify them to reflect their own custom requirements.

To provide QoS in traffic routing using SDN and Northbound Interface, Software-Defined Constrained programming Routing (SCOR) \cite{SCOR} is another solution which is based on Constrained Programming (CP). SCOR divides NBI in two layers as: the upper layer is  \textit{CP Based Programming Language}, and lower layer is \textit{QoS Routing and Traffic Engineering Interface}. The lower layer is defined to address the requirements and has nine predicates: i.e. network path, capacity guarantee, delay, path cost, etc.  SCOR is implemented in MiniZinc\cite{MiniZinc} which is a declarative constrained programming.

{\bfseries{Conclusion:}}
SDN languages are evolving to enhance the abstractions for programmability in networks. There are a number of SDN languages which can take advantages from some new features of OpenFlow. However, it requires adding new libraries and active support from research community. Currently, SDN languages have limited libraries as well as community based contributions, which can be an active area for development.

\subsection{Controller-based and Intent-based NBIs}
Due to absence of a standard northbound interface, most of the available solutions are vendor specific. Some of the controllers provide their own high level interface, whereas some use adhoc APIs. Here, we discuss four controllers for their northbound interfaces. Out of these four controllers, two (Onix and PANE) proposed their own high level API, whereas two most popular controllers (OpenDaylight and ONOS) are using multiple APIs for different northbound functionalities.

{\bfseries Controller-based NBIs:}
Participatory Networking (PANE) \cite{PANE} is an example of SDN controller, which provides its own high level API. This API between control plane and applications allows reading the current state of the network and writing configuration. It involves two major issues; 1) decompose the control and visibility of the network, and 2) resolve the conflicts between different participants. To solve these issues, it uses three types of messages; requests, queries, and hints. Request messages are used for network resources (e.g. bandwidth and access control). A request may effect the state of network for a time interval. Queries are used to read the network state (e.g. traffic between hosts and bandwidth available). Hints provide the network information which may help to improve the network performance. Moreover, it provides an API where end host applications can dynamically request network resources (e.g. bandwidth reservation). To avoid starvation and exceeding the bandwidth limits set by administrator, it uses a verification engine. Although very useful, the effect of excessive requests is not addressed in this work.

Onix \cite{Onix} is another example of controller which provide its own northbound interface. It defines a general API which enables scalable application development. It also allows control applications to read and write the state of network elements. Moreover, it uses a data centric approach which provides consistency between control applications and underlying network devices. It consists of a data model, representing network infrastructure, where each network element corresponds to one or more data objects. Control logic reads the current state associated with a particular object, operates on this object to alter the state, and registers notifications for state changes on this object. A copy of these notifications and changes are also placed in Network Information Base (NIB). Detailed discussion on NIB is given in Section VI. 

ONOS \cite{ONOS} is one of the most popular open source SDN controllers which is driven by OpenNetworks Laboratory (ON-Lab) founded in 2012. ONOS mainly focuses on scalability, high performance, resilience, and next generation device support. It uses a collection of Open Services Gateway initiative (OSGi) bundles and provides interaction with applications by using Java and REST APIs. It supports both command line and graphical user interface to provide flexibility in application development and network administration through REST API. It also provides a wide range of templates for the development of new applications. Due to distributed nature of this controller, it uses general-purpose Remote Procedure Call (gRPC) \cite{gRPC} which simplifies creation of distributed applications. In gRPC, methods of a client application can be called directly on server application, running on a different machine, as it was a local object.

Another open source project is OpenDaylight \cite{ODL}, founding member of Linux Foundation Networking (LFN), which is widely supported by industry and research community. This controller project was started in 2013 and written in Java with the focus on network programmability. ODL also uses OSGi bundles which run as Apache Karaf \cite{karaf} components. DLUX \cite{DLUX} in ODL is used as a web based interface and represents a number of features including graphical user interface for topology representation. Most of the interfaces can be visualized through Yang-User Interface \cite{YangUI}. Yang-UI is collection of REST APIs which enables developers to query network information as well as configure it. For example, network topology component in Yang-UI provides comprehensive information of whole network, and inventory component provides a detailed information of statistics. 

\begin{table}[!t]
	\centering
	\caption{Controller based and intent based NBIs}
	\setlength\tabcolsep{2.5pt}
	\begin{tabularx}{\linewidth}{|>{\RaggedRight\hsize=0.9\hsize}X|>{\RaggedRight\hsize=1.4\hsize}X|>{\centering\hsize=1.2\hsize}X|>{\centering\hsize=0.7\hsize}X|>{\hsize=0.8\hsize}Z|}
		\hline
	    \textbf{Controller/ Literature}	& \textbf{NBI} & \textbf{Controller or Intent based}	& \textbf{GUI}	& \textbf{Security}\\ \hline
		
		ODL~\cite{ODL}	& OSGi and REST APIs	& Both	& Yes	& Strong	\\ \hline
		Onix~\cite{Onix}	& Onix API	& Controller	& No	& Average	\\ \hline
		PANE~\cite{PANE}	& PANE API	& Controller	& No	& Weak		\\ \hline
		ONOS~\cite{ONOS}	& Java and REST APIs	& Both	& Yes	& Strong	\\ \hline
		Pham~et~al. \cite{Pham}	& Programming APIs, CLI and REST API	& Intent	& Yes	& Weak	\\ \hline
	\end{tabularx}
	\label{tab:9a}%
\end{table}

{\bfseries Intent-based NBIs:} 
Adoption of SDN critically depends on its ability to support multiple types of applications through NBI. Most of the solutions for NBI are adhoc, vendor specific and have limited capabilities. Intent based model attempts to resolve these issues and allows declaration of high level policies, instead of detailed specification of different networking mechanisms. The above-mentioned ONOS and ODL controllers also support intent-based northbound interfaces.

Intent framework of ONOS allows applications to specify their requirement of network control in the form of policies instead of mechanisms \cite{IntentONOS}. These policy based directives are referred as Intents. These high level intents are translated into installable forwarding rules which are essential operations to control network. Moreover, these intents can be identified by using two parameters; Application\_ID and Intent\_ID. Application\_ID represents an application which create a particular intent. Whereas, Intent\_ID is generated whenever an intent is created. As soon as an intent is submitted by an application, it is directly sent to the compilation phase. This phase uses an intent compiler, which converts them into installable intents. If an application asks for an unavailable objective (e.g. connectivity among non-connected segments), this phase will see for an alternate approach to recompile. After compiling an intent, it is sent to installing phase, where an intent installer is responsible to convert installable intents into flow rules. An intent manager provides coordination between intent provider and intent installer. 

Network Intent Composition (NIC) \cite{IntentODL} is an internal project of OpenDaylight which is currently in incubation state. NIC provides an interaction between core modules of ODL or external applications to fulfill user desires. It uses current network service functions of OpenDaylight and southbound interfaces to control virtual and physical network elements. In ODL, a component referred to as \textit{renderer} is used to transform the intents to the implementation of flow rules. A wide range of renderers are supported in various versions of ODL, which includes; Network Modeling (NEMO) renderer, OpenFlow renderer, Virtual Tenant Network (VTN) renderer, and Group-Based Policy (GBP) renderer. There are two core functions (i.e. hazelcast and MD-SAL) which supply the base models for NIC capability. On top of these functions, a renderer can be installed. This renderer transforms an intent using a particular project (e.g. VTN, NEMO, etc.) for network modifications. For example, NEMO renderer is a feature that will transform an intent to a network modification by using NEMO project \cite{ODLrenderer} in ODL. 

To create new services as well as to compose or split current services of network applications, Pham et al. \cite{Pham} proposed a solution for intent based NBI. The design principles of this study are: data decentralization, web service components, process isolation, and robustness. Moreover, it proposed a three tier architecture. To provide flexibility, these tiers work independently (i.e. every tier can change its components without effecting other tiers). The tiers are; database tier, business logic tier, and presentation tier. Application states are stored in database tier, which can be retrieved in different contexts. Service creation and composition is handled by business tier. In this tier, a service registry is used to discover existing services whereas, new services can be created as atomic service. To integrate new and existing services, service integration element is used. By using CLI, REST, and programming interface, presentation tier takes input from user. To analyze the process it uses Domain Driven Design (DDD) where requirements are decomposed into smaller problems and then solution for each problem is built. For example composite intents are decomposed into based intents which are further decomposed into solution intents.

{\bfseries{Conclusion:}} 
Due to a diverse range of controller based NBIs, applications designed for one controller may not work for any other controller. However, it is a good initiative to have intent based NBIs but only a few controllers support it. Table~\ref{tab:9a} presents a summary of different controller based and intent based NBIs, where ODL and ONOS supports both controller based and intent based NBIs. PANE and Onix offer their own NBIs. Unlike southbound interface, a mature and comprehensive solution for NBIs is still missing.

\begin{figure*}[!h]
	\centering
	\includegraphics[width=\linewidth]{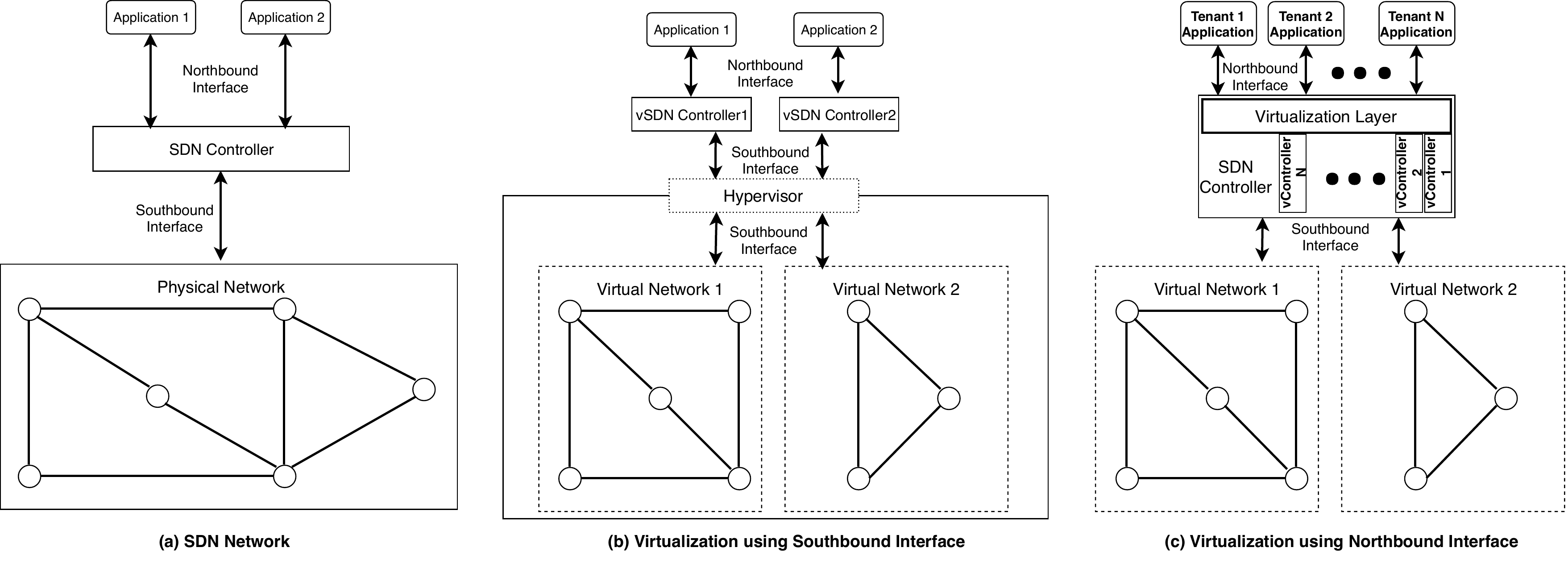}
	\caption{SDN interfaces and virtualization.}
	\label{fig:10}
\end{figure*}

\section{Virtualization and SDN Interfaces}
Current networks are based on devices which are purpose built which directly impacts network up-gradation \cite{NFVNBI}. Whenever a new service is required, new hardware has to be deployed, installed, and connected by some order. It leads to time consumption, complexity, cost, and threats of being error prone.  Network Function Virtualization combined with SDN is a new paradigm which allows noticeable growth. These two technologies minimizes the cost, maximizes network resource utilization, and at the same time reduce complexity of manual configurations \cite{NFVSurvey}.

Fig. \ref{fig:10} represents the relationship of SDN interfaces and virtualization. Conventional SDN architecture is depicted in Fig. \ref{fig:10}(a), whereas a hypervisor is placed at southbound interface as shown in Fig. \ref{fig:10}(b). In this scenario, physical network is divided into multiple virtual networks to make slices, and different applications can run on virtual SDN controller. Sometimes tenants do not require a full fledge controller. In this situation, virtualization schemes run as a module on controller as shown in Fig. \ref{fig:10}(c). Thus, different tenants can run their applications on a single controller using northbound interfaces.

\begin{table*}[]
	\centering
	\caption{Summary of approaches used for virtualization}
	\setlength\tabcolsep{3pt}
	\begin{tabularx}{\linewidth}{|>{\RaggedRight\hsize=0.8\hsize}X|>{\RaggedRight\hsize=1.3\hsize}X|>{\centering\hsize=0.6\hsize}X|>{\centering\hsize=0.6\hsize}X|>{\centering\hsize=0.7\hsize}X|>{\RaggedRight\hsize=2\hsize}X|}
		\hline
		\textbf{Literature}	&{\textbf{Placement}}	& {\textbf{Tenant's Controller}}	&{\textbf{APIs Involved}}	&{\textbf{SBI Support}}	&\textbf{Description}	\\ \hline
		
		FlowVisor~\cite{NFVNBI1}	&Between Switch and Controller	&M	&SBI	&OF	&
		To provide an isolating between experimental traffic and data traffic	\\ \hline
		
		AutoSlice~\cite{AutoSlice}	&Between Switch and Controller	&M	&SBI	&OF	&
		Improve scalability by handling large number of flow tables				\\ \hline
		
		OpenVirteX~\cite{OpenVirteX}	&Between Switch and Controller	&M	&SBI	&OF	&
		Provides Topology, Address and Control Function Virtualization			\\ \hline
		
		FlowN\cite{FlowN}	&Inside Controller	&S	&NBI	&OF	&
		Allows tenants to run their own applications rather than having full control of network		\\ \hline
		
		Network Hypervisor \cite{NetworkHypervisor} &Inside Controller	&S	&NBI	&Multi Technology	&
		Handles seamlessly different level of abstractions and a variety of APIs in SDN				\\ \hline
		
		NVP~\cite{NVP}	&Inside Controller	&C	&NBI	&OVS	&
		Allows cloud tenants to manage their data center resources				\\ \hline
		
		AutoVFlow~\cite{AutoVFlow}	&Between Switch and Controller	&M	&SBI	&OF	&
		Provides Complete "Flow Space" Virtualization in WAN					\\ \hline
		
		libNetVirt~\cite{libNetVirt}	&Inside Controller	&S	&NBI	&Multi Technology	&
		Provides a flexible way to create and manage virtual networks			\\ \hline
		
		\multicolumn{6}{>{\centering\setlength{\hsize}{6\hsize}\addtolength{\hsize}{10.95\tabcolsep}}X}{Legend: M= Multiple, S=Single, C=Cluster, OF=OpenFlow, OVS=OpenVSwitch}
	\end{tabularx}
	\label{tab:9}%
\end{table*}

OpenFlow networks have the potential to open the control of network but only one user can work on network devices at a time. FlowVisor \cite{NFVNBI1,NFVNBI2} allows multiple users to operate independently on their slices without any conflicts. FlowVisor acts as transparent proxy where OpenFlow messages generated by network devices go to the FlowVisor from where they are routed to appropriate users. In this way, FlowVisor acts as a virtual controller for underlying switches and virtual switch for the users. It partitions the flow tables in so called flow-spaces and in result OpenFlow switches can be manipulated by multiple controllers. Furthermore, Auxiliary Software Datapath is applied to overcome the limitation of flow table size in OpenFlow switches.

AutoSlice \cite{AutoSlice} proposed another solution of virtualization of underlying network devices with the main focus on scalability. It uses hypervisor, placed between switch and controller, which can handle large number of flow table control messages and multiple tenants. This hypervisor is composed on Management Module (MM) and multiple Control Proxies (CPX) which can evenly distribute the control load. Upon receiving a request, Management Module determines the appropriate Virtual SDN (vSDN) it belongs to, and CPX install flow entries accordingly.

OpenVirteX \cite{OpenVirteX} is an approach which provides topology virtualization, address virtualization, and control function virtualization. Tenants can request a topology by providing a mapping between the elements in physical topology and desired virtual topology. This virtual topology can either be an exact physical topology or a subgraph of it. It also grants permission to tenants for custom address assignments. Multiple addresses can cause problems at the time of flow installation. To resolve this problem OpenVirteX generates a globally unique tenant ID, to allow every tenant to run its own network operating system which maps various control functions for the virtual network.

Another virtualization scheme for WAN is AutoVFlow \cite{AutoVFlow} which uses multiple controllers. It resides between switch and controllers and uses southbound interface for virtualization. It allows delegation of configuration role to multiple administrators. It implements a mechanism of flow space virtualization on Wide-Area Network without any need of third party software as it is done in OpenVirteX \cite{OpenVirteX}.

In some cases, a tenant does not require a full fledge control over the network. In such a situation, each tenant can use FlowN \cite{FlowN} which resides inside a controller and uses northbound interface. It allows tenants to run their own applications over a single SDN controller. Authors in \cite{FlowN} used NOX as SDN controller to implement FlowN. Rather than running controller for each tenant, they used shared controller for all the tenants. Virtualization in FlowN is container based where applications running on top of controller consists of handlers which respond to network events. Each of these applications have the illusion that they are running on their own controller rather than a shared one.

Network Hypervisor \cite{NetworkHypervisor} seamlessly handles complexity of different level of abstractions and a variety of APIs in SDN. Similar to FlowN, it also acts as a controller to the applications and provides visualization of the underlying network devices. By using this approach, SDN applications interact with Network Hypervisor through northbound interface and compiles the attributes of respective APIs. It is implemented on top GENI testbed and supports GENI API \cite{GENIAPI}.

Network Virtualization Platform (NVP) \cite{NVP} uses Onix \cite{Onix} controller platform where cloud tenants can manage data center network resources by using OpenVSwitch (OVS). Rather than tenants running their own controller, it provides a virtual slice to tenants to manage their resources by running their applications through high level API. It resides inside the controller and uses northbound interface for virtualization purpose.

libNetVirt \cite{libNetVirt} is an approach which is used for virtualization of networks in the same way as it can be done in machine virtualization. It is divided into two components; i) generic interface, and ii) drivers. Generic Interface is a set of function that allows the interaction between virtual networks. Drivers, on the other hand, are the technology dependent elements. For this virtualization scheme northbound interface is involved. For southbound interface it is not dependent on OpenFlow, and other proposal can also be used.

Table \ref{tab:9} presents a summery of virtualization techniques along with the details of interfaces involved. Some proposals use hypervisor which is placed and works as a proxy between switches and controllers. Whereas, in some cases a full fledge controller is not required by tenants. In this case, controller is divided into multiple virtual controllers and tenants can run their application by using their own slice of virtual controller.

\section{East/Westbound Interface (E/WBIs) in SDN}
The main advantage of SDN is to provide a centralized view of the network. But exponential increase in network devices has led to new challenges. On one hand, deploying a new SDN domain is a relatively simple approach, but to make this new domain interoperable with traditional Autonomous Systems (AS), is challenging. A common method for this purpose is to use BGP for information sharing. However, PCEP \cite{PCEP} and GMPLS \cite{GMPLS} can also be used for communication among SDN controllers and legacy networks \cite{6994333}. However, these solutions are not designed for SDN, rather they are used as makeshift solutions. East/Westbound Interface is used for interconnection between different SDN domains or interaction between SDN and traditional network domains, where east refers SDN-SDN communication, and west refers to legacy-SDN communication.

\subsection{Interaction between SDN Domains (Eastbound APIs)}
Development in SDN provides an opportunity for networking innovations. Centralized control makes network programmability and network management simple but for large scale networks there are new challenges like scalability, security and availability~\cite{WIBOWO201732}. 
\begin{itemize}
\item {\bfseries Scalability:} A single controller can manage only a limited number of switches, which causes scalability issue. In SDN architecture, a centralized controller is the key artifact, but it also creates a performance bottleneck as soon as number of switches increases. It also introduces a risk of single node failure problem.
\item {\bfseries Security:} A range of security problems may effect SDN controller and its performance. In case of a large scale network \cite{DoS}, a Denial of Service (DoS) attack may lead to a worst case scenario.
\item {\bfseries Availability:} Every time a new packet arrives, data plane devices need controller's involvement to process that packet. Overloaded controller may not be available for devices at all times. For example NOX \cite{NOX} can handle 30K flow requests with a response time of less than 10ms, but for larger networks, it may be higher \cite{Tootoonchian:2012:CPS:2228283.2228297,KARAKUS2017279}.
\end{itemize}

\begin{figure*}
	\centering
	\includegraphics[width=\linewidth]{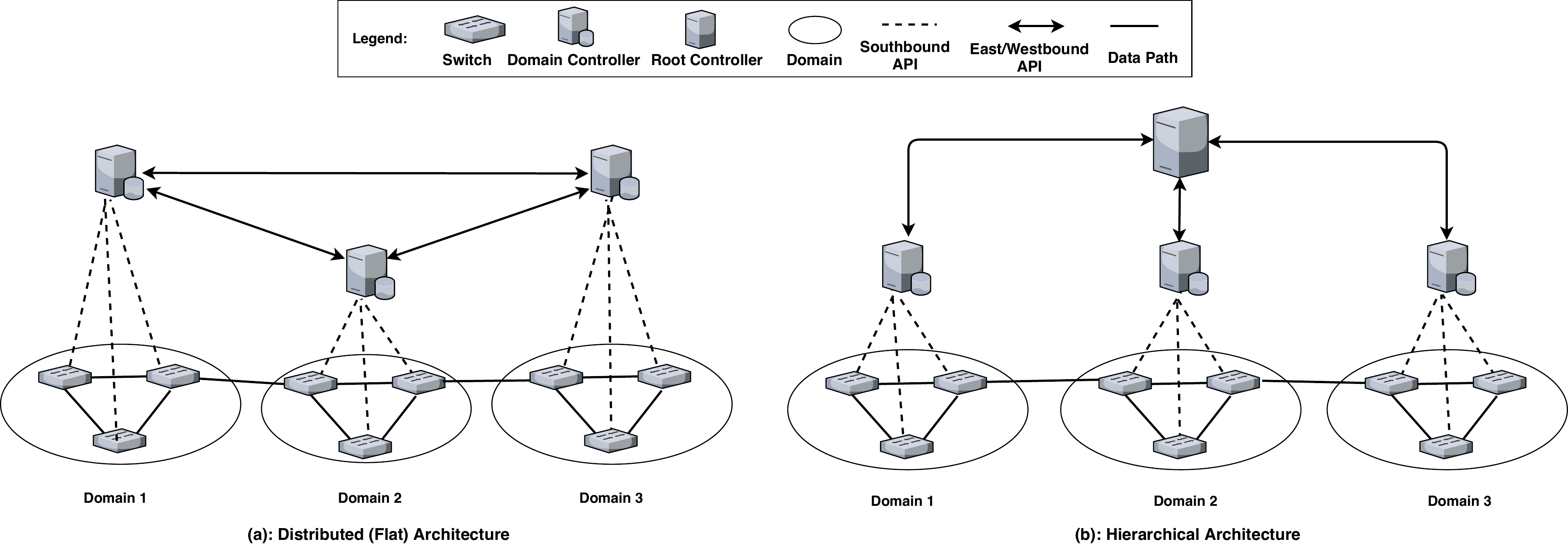}
	\caption{Classification of distributed SDN architectures.}
	\label{fig:11}
\end{figure*}

Data Center Networks (DCNs) are prime candidates where SDN is implemented, but due to its features enterprise level networks have also adopted SDN. Implementation of SDN at enterprise level is difficult due to issues discussed above. A simple solution is distribution of SDN controllers. Such distributed controllers can be implemented as: 1) Distributed (Flat) Architecture 2) Hierarchical Architecture. In distributed architecture, all the controllers have equal rights and share information (e.g. topology, reachability, devices capabilities, etc.) with each other, as shown in Fig. \ref{fig:11}(a). On the other hand, hierarchical architecture has two layers of controllers. Lower layer consists of domain controllers, sometimes referred as local controller, and upper layer contains a root controller. Local controllers are responsible for their own domain and update root controller by using a control channel, as shown in Fig. \ref{fig:11}(b). Root controller normally has more rights as compared to domain controllers and keeps network wide information.

Table \ref{tab:10} presents a summary of these architectures along with the list of protocols used, their network types and languages used by different proposals. Some of the proposed architectures are using distributed approaches, some of the proposals are with hierarchical and proposals like FlowBroker and Orion are using a mixture of these two approaches.

\subsubsection{\bfseries Distributed Architecture Interfaces}
This architecture can be classified into two types. One is logically centralized but physically distributed and other is completely distributed. In logically centralized but physically distributed SDN architecture, each controller is responsible for its own domain but synchronized with other controllers. As soon as there is any change under any controller, it will update neighboring controllers. This enforces a consistent global view of the network. A key problem with this approach is that controllers consume network resources to provide information to each other and frequent change in network may reduce network performance. 
On the other hand, in completely distributed controller architecture, controllers are not synchronized. They may update each other using protocols, but consistency does become an issue in this approach, which may lead to unexpected behavior of network.

SDNi \cite{SDNi} attempts for inter-controller communication applied in ODL \cite{ODL}. It is a message exchange protocol among different domains coming under single operator or collaborating operators. It exchanges customized messages like reachability, flow setup, and capability updates.

HyperFlow \cite{Hyperflow}, an event based solution, uses WheelFS \cite{WheelFS} as a file system for controller communication in a domain. It is a logically centralized and physically distributed architecture where switches connect to the nearest controller which updates neighboring controllers by using publish/subscribe method. It provides a consistent global view of the network. HyperFlow runs as an application on top of NOX \cite{NOX} controller and uses most of the features of NOX.

Onix \cite{Onix} is another approach to overcome scalability problem in SDN controllers, and provides flexibility for the development of applications in distributed manner, by providing Distributed Hash Table (DHT) storage and group membership. It has three major components; (1) Network Information Base (NIB), (2) Partition and Cluster Aggregation for hierarchical structure, and (3) Consistency and Durability for applications. NIB is a data structure that maintains network entities. To access any particular entity, it queries the index of all entities by using entity identifier. Moreover, NIB can cause scalability issues (e.g. exhaust system memory and saturate CPU or Onix instances) as it is not distributed. To resolve this issue, it uses partition and cluster aggregation. Control applications in Onix are used to partition the workload. Whereas in aggregation, cluster managed by different Onix nodes is considered as a single node. Moreover, authors claim that consistency and durability can be achieved by using different algorithms, however details for these algorithms are missing.

Tam et al. \cite{5928883} used two approaches to resolve problem of scalability and multi path among different controllers without using global view in data center environment. It uses multiple independent controllers to answer the request of underlying devices, instead of a single omniscient controller. First approach is Path-Partition Approach; where all possible paths are calculated from source to destination using Dijkstra's Algorithm and then each multi path is allocated to one of the controllers according to cost function and number of links monitored by that controller. Whereas, Partition-Path approach uses initial preferred links for path computation from source to destination. All the controllers have different routes of source and destination pairs. A source can send request to all the controllers for route request. This solution will generate extra control traffic. Another solution is to have a  mapping at source node where a table at source node describes which controller has route for particular destination.

Switch to controller mapping is static and a controller may become overloaded if large number of flows arrive. It is quite possible that other controllers might be underutilized because of static mapping, which may lead to poor performance. Elasticon \cite{Elasticon} proposes an architecture, where load of a controller is computed and controller pool can be dynamically expanded and shrunk, which enhances the network performance and throughput. A switch can be connected to multiple controllers, with one master and rest as slaves, for fault tolerance purpose. A distributed data store is used to provide communication among different controllers and enables logically centralized controller. It also has switch-specific information and each controller maintains a TCP connection with other controllers in the form of mesh. This TCP connection is used to send messages to coordinate with other controller while switch migration. This connection is also used to send messages to a switch which is connected with other controller as master.

ONOS \cite{ONOS} also provides an approach for improving scalability and fault tolerance to SDN control plane. It first of all creates a global network view by using Titan \cite{Titan} graph database with Cassandra \cite{Cassendra} key-value store for distribution. There are multiple instances but only one instance is the master for each switch, which is responsible for taking information from switch and program it. One of the issue in first ONOS prototype was the implementation of notifications and messaging across the ONOS instances. For changes in network states, ONOS modules had to check the database periodically which increases the CPU load as well as delay for reacting to events and communication among instances. This issue was resolved in second prototype by creating an inter-instance communication module using publish-subscribe mechanism based on Hazelcast \cite{Hazelcast}.

To facilitate the deployment of SDN in large scale and to do traffic management by the coordination  Helebrandt et al. \cite{7107574} proposed an architecture for communication among multiple domains using an interface, which is referred as INT module. The protocol is divided in three sub-parts; 1) Controller Interconnection Session Control, 2) Capabilities Information Exchange, and 3) Path Setup. Controller interconnection session control is responsible for connection establishment. To reduce the administrative overhead, this session can be automated, but for security purposes it has manual setup. Capabilities Information Exchange are used to exchange the capabilities of network elements, whereas path setup is used for end to end flow setup. Another responsibility of this protocol is to send keep alive messages and provide updates to peer controllers. Furthermore, four types of messages are used for this purpose which are request, reply, help, and update. A sample packet header is also discussed for communication among multiple domains.

In \cite{DBLP:journals/corr/YaziciSE14}, Yazici et al. proposed a framework which provides support for dynamic addition and removal of controllers to the cluster without any interruption, and at the same time the framework can work with numerous existing SDN controllers. It is a leader based approach where JGroup \cite{JGroup} notification and messaging infrastructure is used for the communication among different controllers to select master controller. Master controller is responsible for delineation between different controllers and switches. If for any reason master controller is not accessible, new master is selected. This architecture is not hierarchical as it allocates some additional responsibilities to master controller.

\begin{table*}[]
	\centering
	\caption{Summary for inter controller communication interfaces (EBIs)}
    \setlength\tabcolsep{3pt}
    \begin{tabularx}{\linewidth}{|>{\RaggedRight\hsize=0.8\hsize}X|>{\centering\hsize=0.6\hsize}X|>{\RaggedRight\hsize=0.8\hsize}X|>{\RaggedRight\hsize=0.8\hsize}X|>{\centering\hsize=0.9\hsize}X|>{\RaggedRight\hsize=2.1\hsize}X|}
    	\hline
    	
	    \textbf{Literature} & \textbf{Architecture} & \textbf{Protocol for Communication} & \textbf{Network Type} & \textbf{Prog. Language Used} & \textbf{Description} \\\hline
	    
	    ODL \cite{ODL,SDNi}	&Distributed	&SDNi Using BGP	&WAN	&JAVA	&SDN interface (SDNi) enables controller to exchange information within the purview of define policies 	\\ \hline
	    
	    Kandoo \cite{Kandoo} & Hierarchical & Messaging Channel & Data Center, Campus & C/C++/Python & Divides control plane in domain controller and root controller \\
	    \hline
	    
	    DISCO \cite{DISCO} & Distributed & AMQP  & Data Center, Enterprise, WAN & JAVA & Based on Messenger and Agent Approach which are responsible for control information \\
		\hline
		
	    Onix \cite{Onix} & Distributed & Zookeeper & Data Center, Enterprise & C++   & Provides an API for the easiness in application development \\
	    \hline
	
	    HyperFlow \cite{Hyperflow}	&Distributed	&WheelFS	& Data Center &C++	& An event based solution running on top of NOX and provide a consistent global view of the network \\ \hline
	    
	    Tam~et~al. \cite{5928883}. & Distributed & Not Mentioned & Data Center & Controller Dependent & Allow controllers to distribute their loads to reduce response time in Data Center Environment \\ \hline
	    
	    Elasticon \cite{Elasticon} & Distributed & Custom Protocol & Data Center, Cloud & JAVA  & Ensures controller utilization by computing controller load and stretch or shrinks accordingly \\ \hline
	    
	    ONOS \cite{ONOS} & Distributed & Raft & Enterprise, WAN & JAVA & Provides scalability and fault tolerance in control lane by using instance based approach \\ \hline
	    
	    Helebrandt~et~al. \cite{7107574} & Distributed & Custom Protocol & WAN   & Not Mentioned     & Enables Controller Communication using INT module responsible for connection establishment and keep alive messages \\ \hline
	    
	    Yazici~et~al. \cite{DBLP:journals/corr/YaziciSE14} & Distributed & JGroup & Data Center & JAVA  & A Master controller is selected among different controllers by using Jgroup and a controller can added or removed without network interruption \\ \hline
	    
	    WE~Bridge \cite{WEBridge} & Distributed & Custom BGP & Enterprise, WAN & JAVA & Provides Scalability and control messages are forwarded in JSON format \\ \hline
	    
	    DMC \cite{chundrigar2016dmc} & Distributed & RabbitMQ & Data Center, Enterprise, WAN & Python & Ensures flexibility in link and controller failure among heterogeneous domains \\ \hline
	    
	    Bari~et~al. \cite{Bari} & Distributed & Not Mentioned & WAN   & Python & Solving Dynamic Controller Provisioning Problem and reduce flow setup time. \\ \hline
	    
	    Orion \cite{Orion} & Distributed \& Hierarchical & Not Mentioned & WAN & JAVA & Solves path stretch problem and super linear complexity by combining distributed and hierarchal architecture \\ \hline
	    
	    Bhole~et~al. \cite{7546147} & Hierarchical & Not Mentioned & WAN & JAVA & Improve communication reliability and reduce response time \\ \hline
    
	    FlowBroker \cite{Marconett2015} & Distributed \& Hierarchical & Broker Protocol & WAN   & JAVA  & Improves load balancing and network performance in multiple domains  \\ \hline
	    
	    Karakus~et~al. \cite{Karakus}& Hierarchical & Broker Protocol & WAN   & Controller Dependent & Providing Quality of Service (Qos) using FlowBroker Approach \\ \hline
	    
	    Guo~et~al.\cite{7446851} & Hierarchical & NBI & WAN & JAVA  & An hierarchical architecture using Northbound API for communication among different controllers \\ \hline
	    
	    Wang~et~al. \cite{7864285} & Hierarchical & Restful API &  Enterprise, WAN &Not Mentioned  & Co-ordinate controller is used to provide communication among heterogeneous controllers \\ \hline
	    
	    D-SDN \cite{DSDN} & Hierarchical & Custom Protocol & Home, WAN & Not Mentioned & Using Master and Secondary approach where master controller is delegating control to secondary \\ \hline
    \end{tabularx}%
  \label{tab:10}%
\end{table*}%

In DISCO \cite{DISCO} authors used FloodLight OpenFlow Controllers in multiple domains. The solution provides Intra-Domain and Inter-Domain Communication and is resilient from disruptions. DISCO's architecture is mainly divided into two parts: Messenger and Agent. Messenger is responsible for light weight control communication among different controllers by using AMQP \cite{AMQP} protocol using publish-subscribe method, whereas, Agent supports network wide functionalities like Connectivity, Monitoring, Reachability, and Reservation. Furthermore, agents identify alternative routes to offload traffic from weak interconnections. If they can not find any alternate routes, they reduces the frequency of control messages for these weak interconnections.

Implementation of WE Bridge \cite{WEBridge} uses heterogeneous controllers and provides a bridge for communication among these controllers. It first registers controllers and then provides virtualization among domains because it is possible that Internet domains may belong to different administrative authorities. Domains are using an interface to transfer messages in the JSON format whereas other options are also recommended like; XML and Yanc. Two different applications Inter Domain Path Computation and Source-Address based Multi path Routing run on top of controllers.

Distributed Multi-Domain Controller (DMC) \cite{chundrigar2016dmc} connects heterogeneous networks. This provides privacy of domains and at the same time deals with link and controller failure among different domains. Light-weight controller-to-controller communication is done by using RabbitMQ \cite{RabbitMQ} which is an implementation of AMQP. Controller in each domain is event driven and provides services in its own domain and communicates with neighboring domains by using control channel which is integrated with REST interface. A centralized database is managed where all the controllers update data. Publish-Subscribe method is used among controllers.
 
\subsubsection{\bfseries Hierarchical Architecture Interfaces}
In hierarchically distributed architecture there are two layers of SDN controllers. Lower layer controllers are responsible for their respective domains. At upper layer, root controller is responsible for managing a group of domain controllers. Control information in this architecture is less as compared to flat architecture but Single Node Failure problem still exists, however it is not as problematic as earlier.

To make SDN scalable, number of frequent events (e.g. network wide statistics collection and flow arrivals) in control plane must be reduced. It can be done by processing these events in data plane which is a costly solution, as it requires switch modifications. A solution of this problem is addressed by Kandoo \cite{Kandoo} which is a hierarchical control plane architecture with two layers of controllers. Lower layer of controller deals with their own domains and does not have a network wide view. Moreover, controllers in this layer are subjected to deal with frequent events. Whereas, top layer which consists of a root controller and has the global network state and processes rare events (e.g. elephant flows). It uses an API in terms of two applications as $ \ {App_{detect}}\ $and$ \ {App_{reroute}}\ $. $\ {App_{detect}}\ $runs on local controllers and constantly queries each switch to detect elephant flows. $ \ {App_{reroute}}\ $runs on root controller and install flow entries on switches if elephant flow is detected by $ \ {App_{detect}}\ $.  To differentiate the applications running on local or root controller, a flag is used.

Karakus et al. \cite{7097964} also proposed an architecture with two levels which can be extended. Bottom level is referred as network level, and contains different SDN domains handled by local controllers. Broker level is an upper level where a super controller is placed which supervises domain controllers. Local controllers advertise all their reachable addresses as well as border switches connecting their neighboring domains by using eastbound interface, which is file based, to super controller. It allows super controller or broker to determine the source and destination domains. Whenever there is inter domain communication, broker asks source and destination domains to advertise all QoS paths from source to gateways (in case of source domain) and destination to gateways (in case of destination domain). As broker has the global view of the network, it determines all paths from source to destination. After computing best route, broker sends ingress and egress node points to respective domains (i.e. source, destination, and transit) to reserve the QoS values. Authors also claim that a controller in a hierarchical setting handles 50\%  less number of traffic than a controller  in  a  non-hierarchic environment.

Guo et al. \cite{7446851} used hierarchical model for multi domain controller communication, where local controllers are responsible for their own domains whereas coordinating controller is responsible for the global view of the network and provides inter controller communication which is prototyped in Java. Communication among coordinate controller and domain controller as well as with the applications is done by using Northbound API. This NBI can provide information of local controller to applications and coordinating controller. It also enables applications and Coordinate controller to configure flow tables and traffic forwarding.  Furthermore, two modules are implemented as Topology Management and Flow Management. Topology Management is responsible to get the whole topology and flow management is about updating and installing flows to domain controllers and data plane respectively.

To solve the issue of consistency among diversified controllers, Wang et al. \cite{7864285} proposed a coordinate controller approach which helps for communication among heterogeneous controllers. Control plane in this architecture is divided into two parts, coordinate controller and domain controller. Coordinate controller is responsible for the collection of information of whole network and domain controllers and dynamic controllers may use different technologies. Domain controller is a traditional SDN controller and running its own domain. Protocol interpreter is used for eastbound communication among coordinate controller and domain controller which enables coordinate controller to implement end-to-end provisioning services across multiple domains. To resolve the issue of diversity of vendors, it uses a unified Northbound interface. This unified API is divided into two parts; topology API and service API. Topology API is used for the collection of network information and elements connectivity to design a global view of the network. Purpose of service API is to launch service requests and setup a connection in the network.

Sometimes control can also be delegated to underlying devices. For example Decentralized SDN (D-SDN) \cite{DSDN} is a hybrid approach using Main Controller (MC) and Secondary Controller (SC) by delegating control. It allows physical as well as logical control distribution by using MC and SC. One of the integral feature in D-SDN is security, as MC authorizes before delegating control to SC so that it can act as controller. This delegation occurs upon a request from SC that is triggered by a set of events. These events include a newly installed SC, or a gateway through which mobile devices can access Internet. SC can not write any new flow entries without authentication of MC.  Communication between MC and SC is done by using an interface for control delegation message where SC requests a Check\_in\_Request and whereas master authorizes or denies Check\_in\_Response. Similarly, communication among SCs is done by using D-SDN's SC-SC protocol to implement fault tolerance. Switch to controller mapping in SCs are master-slave based. A slave SC can receive Hello messages for a predefined period of time. If it does not receive, slave SC can become master by sending role\_change message to the switch.

Every controller has different features and  northbound interfaces, for example, deleting a flow in Floodlight controller is easier as compared to POX and both of these controllers have different functions.  
Zebra \cite{Zebra} attempts to resolve this heterogeneity problem by dividing control layer in two parts; dissemination layer and decision layer. Dissemination Layer has traditional SDN controllers, whereas, two main modules referred to as; Heterogeneous Controller Management (HCM) and Domain Relationships Management (DRM) are placed in the decision layer. Different SDN controllers (e.g. FLoodlight, POX, OpenDaylight etc) are placed in HCM and it handles the routing decision inside a domain.  Whereas, CRM provides decision making among multiple domains.

\subsubsection{\bfseries Hybrid Architecture Interfaces}
Orion \cite{Orion} is an example for large scale networks with a mixture of distributed (flat) and hierarchical architectures. It mainly focuses on Path Stretch problem and Super Linear Computation Complexity problems introduced by these two architectures. Path stretch is the difference between best optimal path and actual path traffic takes in the network. This problem occurs in hierarchical architecture. Super linear computational complexity issue exists in distributed (flat) architecture and normally occurs because of size of network. Orion addresses these issues by dividing architecture in three layers which includes, physical layer, area controller layer, and domain controller layer. Area controllers are close to OpenFlow switches and pass on the information to domain controllers which consider area controllers as nodes and reduce the path stretch and super linear computational complexity problem. A TCP connection is used as an interface for communication purposes between area controller and domain controller. This channel is used for sending requests and distribute rules.

In \cite{Marconett2015}, Marconett et al. proposed a mechanism called FlowBroker using hierarchical approach to do load balancing and improve network performance. Brokers work as root controller on top layer, whereas, in lower layer, domain controllers are managing there own domains. Each domain controller is attached to a broker according to their reputation in terms of load balancing and performance. This performance reputation by using machine learning based agents which are connected with domain controllers. A broker is a software process which allows exchange of network wide state along with the flow table updates to respective domain controllers. To counter the failover mechanism, switch controller mapping is done by using primary and secondary controllers. Secondary controllers monitor primary controller by using an interface based on Ctrl\_Keep\_Alive messages after every two seconds. Whereas, primary controller sends Ctrl\_Table\_Backup messages periodically to mirror any changes occur in primary controller.

{\bfseries{Conclusion:}}
Distributed controller approaches solve a number of issues in SDN, but it also introduces some new challenges, for example, consistency and resource utilization. Similar to northbound interface, east interfaces are also not standardized which is one of the major challenge in distributed controllers. Moreover, a consistent global view is also required in distributed controllers which may reduce network performance. Controlling the overhead is another major challenge. From a wireless network perspective, lightweight controllers for access points could lead to better flow management in mobile networks. Hence, EBIs for inter-AP Controller communication will certainly benefit softwarization of wireless networks.

\subsection{Interaction between SDN and Traditional Networks (Westbound APIs)}
Some critics believe that Inter Domain communication in legacy network is better rather than using SDN. For example in \cite{7878790} authors use BGP on top of TCP for inter-controller communication. Session starts by using OPEN message which leads to ESTABLISHED once connection is established among these controllers. Controllers can share information by using UPDATE messages which includes reachability and messages like bandwidth information. Authors claims that legacy networks are performing well as compared to SDN with BGP and SDN without BGP.

\begin{figure}
	\centering
  \includegraphics[width=\linewidth]{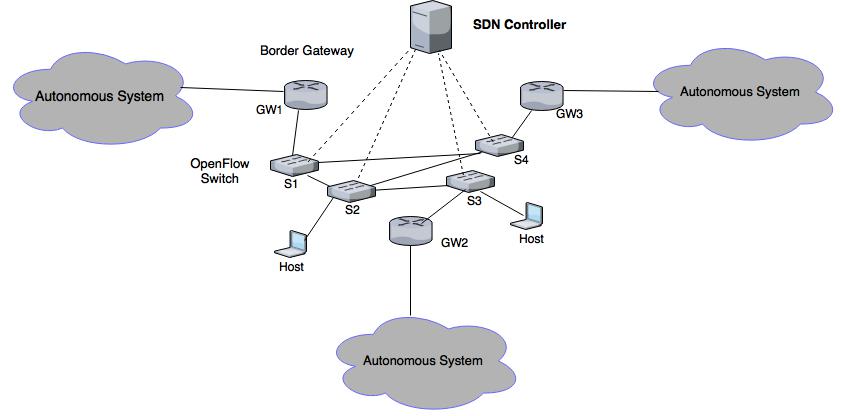}
  \caption{SDN and autonomous systems.}
  \label{fig:12}
\end{figure}

SDN is a promising way to re-architect the Internet and transitioning from traditional network to SDN is an important issue as there are a large number of Autonomous Systems throughout the globe \cite{F2,SDX}. During this transition SDN must co-exist with legacy network and any SDN network should be able to share reachability information, forward traffic and express routing policies with traditional network through gateways in SDN domain. Fig. \ref{fig:12} presents traditional ASes communication with SDN domain through border gateway switches. Migration Work Group \cite{MONF} under ONF \cite{ONF} is working on proposals for the transition from traditional to SDN networks.

RouteFlow \cite{RouteFlow} is one of the first approaches of IP routing on OpenFlow switches which is composed of RouteFlow Server, RouteFlow Slave, and RouteFlow Controller. RouteFlow Controller runs as an application on top of SDN Controller. RouteFlow Server keeps network wide state and core logic resides in it, and manages a virtual network environment to interconnect virtualized IP routing engines e.g. Quagga \cite{Quagga}. For legacy network RouteFlow Slave is used which updates server using a custom interface (i.e. RouteFlow Protocol). The messages of this protocol are either command type or event type. These messages are the subset of OpenFlow protocol along with some other messages (e.g. send updates, accept,/reject VM, RF-Slave configuration, etc.).

Another solution using BGP is SDN-IP \cite{SDNIP} where seamless interconnection between SDN domain and traditional domain is focused. SDN-IP Peering application (having BGP Route Module and Proactive Flow Installer Module) runs on top of Network Operating System. BGP route module synchronizes BGP route updates pushed by BGP process which is ZebOS BGPd \cite{ZebOS} and store them in Route Information Base (RIB) which can scale to 10,000 entries, whereas Proactive Flow Installer uses routes learned through BGP and installs flow entries accordingly. In simple words, controller of SDN domain uses BGP to exchange routing information with neighboring legacy network domains but uses SDN's centralized mode to control local AS's BGP route calculation and installation.

BTSDN \cite{BTSDN} proposes a practical solution by integrating SDN network to the current Internet with BGP \cite{BGP,BGP_Iqbal}. Using BTSDN, SDN and traditional network can co-exist by using Internal BGP (iBGP) and External BGP (eBGP) so that SDN can be incrementally deployed to the Internet and finally replace traditional networks. Usage of eBGP and iBGP in BTSDN is same as traditional network. OpenFlow switches directly connected to border routers and play a key role and act as a proxy because SDN controller can not directly control border routers. However controller can install certain flow entries on these SDN switches. Data plane adopts the mechanism of Address Resolution Protocol (ARP) and Media Access Control (MAC) to ensure the delivery of IP packets between SDN and traditional domains.

Internet eXchange Points (IXPs) are  playing an integral role in interconnecting many networks and bringing popular content closer to end users. Routing in traditional network uses BGP which is only on destination IP prefix and networks can not make more fine grained decisions based on the type of application or sender. Similarly routes are learned from direct neighbors so network can not provide proper end to end service and at the same time network can not express inbound and outbound paths. To resolve all these issues a combination of IXPs and SDN makes Software Defined eXchange (SDX) \cite{SDX}. It enables their participants to run novel applications, written in Pyretic \cite{Pyretic}, that control the flow of traffic entering and leaving their border routers. It gives an illusion to each AS has a virtual SDN switch connected to its border router and enables flexible specifications of forwarding policies and at the same time it provides an isolation among different participants.

{\bfseries{Conclusion:}}
Newer versions of OpenFlow provides hybrid solutions, where controllers are able to communicate with SDN elements as well as traditional switches. However, research for westbound interface is required during transition period from traditional networks to SDN. Co-existence interoperability will be a key step for large scale SDNs.   

\section{Future Research Directions}
Software Defined Networks have been a major research area in recent years. However, more effort has been places on controller design. Interfaces on the other hand has received less attention other than OpenFlow. Here we present a number of research directions and possible challenges for each type of interface. 

\subsection{Southbound Interface}
OpenFlow has dominated the SBI, as it has matured rapidly, although other solutions (as discussed in this article) also offer interesting features. OpenFlow has underwent rapid evolution, which has its pros and cons. A long header for matching is used in different OpenFlow versions, which leads to the requirement of more storage for rules and takes more processing time. Although number of other studies exist, for reduction of this is size, not all have been integrated into OpenFlow. This can certainly be a development direction. An optimal mechanism to reduce storage of processing requirements will be beneficial to the overall system in different domains.

ForCES offers a rich a set of features, such as, separation of control and data plane without changing the architecture of traditional networks, and extensibility. These features are still not available in OpenFlow or other southbound interfaces. A possible approach could be to further develop ForCES to become a more elaborate solution, or to manage these features in OpenFlow. Combining all possible features in a single solution may make it too complex and increase its overhead, hence further research is needed to adequately evaluate the performance of both, and then develop on their capabilities for specific types of networks.

Two different modes are used to reduce the latency in flow rule setup: proactive and reactive. In proactive mode, flows are already installed before arrival of packets. However, this solution create unnecessary overhead, but can be useful for critical flows which are delay sensitive. In reactive mode time taken for flow installation is crucial because flow installation is done after packet arrival. This mode is not suitable for time sensitive flows. Research challenges in this regard are multi fold Depending on type of network the solution could be different. As SBI that takes into account the design and usage of networks, types of communicating devices, and their mobility traffic patterns may yield better flow installation time. This is also a lack of performance analysis in term of SBI effect on flow installation by different solutions. This is another area of exploration. 

For a wide deployment of SDN in WSN, a number of issues have already been addressed. However, robustness in case of sink failure is a challenge and open question. As more sensor node deployment is being done in different networks, hence it is important to evaluate the softwarization and hence a unified lightweight SBI. Memory size of sensor nodes are limited to keep a large number of flow rules. Similarly, security and mobility management of different nodes requires further research attention. A constant research in the area of energy management is a necessity to ensure an efficient use of energy resources.

A large number of devices are expected to be connected through Internet because of the power of Internet of Things. SDN can help IoT in different aspects. However, to manage the enormous collection of heterogeneous devices via centralized control, an adequate solution in terms of southbound interface is still missing. The fundamental limitation in OpenFlow design context. Current solutions are only designed to communicate with switches. However, in a multi-hoping ad-hoc nature of IoT these solutions have to effectively extend beyond vSwitch. This again requires newer and efficient design for SBI which can communicate with heterogeneous devices with multi-technology interfaces. A unified SBI multi-technology flow installation, topology management, and configuration will be required in future for large scale IoT systems. 

\subsection{Northbound Interface}
SDN provides incredible opportunities for network operators in terms of network management using a centralized controller. However, due to absence of a standardized API, this has become a challenging task. Moreover, it becomes more challenging and time consuming due to distributed controller environments. Hence, to make SDN a powerful option for network operators, an instinctive API is desired. Due to diverse landscape of network elements and multiple versions of protocols (e.g. OpenFlow) portable application development is very difficult. Thus, a flexible interface is required which is capable of removing this underlying complexity.

A number of new features have been introduced in all OpenFlow versions, and programming languages can take advantages of these features. However, FatTire \cite{FatTire} is the only language for NBI which incorporated group tables introduced in OpenFlow version 1.1. Similarly, there are many other features of OpenFlow which can help in various programming languages, but very limited advantages has been taken by high level languages. Efforts in this regard can lead to potential increase in application development for SDN in different domains. Many of the application programming languages offer libraries and community contributed extensions which make them famous in developer communities. However, SDN programming languages do not provide such an interface or repository. Instead, these languages provide fundamental constructs which enforce developers to write applications from scratch.

Vendor specific and adhoc solutions are major issues of traditional networking which exist in SDN as well as controller based northbound interfaces. Intent based interfaces can be a solution of this issue, however further exploration on how to utilize them effectively is required. 

\subsection{Virtualization}
In SDN, data plane performance is directly dependent on control plane performance. There is significant research to enhance the performance of control plane. However, different tenants of vSDN may also need to specify their control plane demands in addition to requesting the virtual topologies and links. Similarly, tenant may specify demands for OpenFlow message types. For example, a tenant may require a fast processing of FLOW\_MOD messages instead of OpenFlow stats. In current research, defining these control plane and OpenFlow specific demands is not available. This research direction in the long run will allow fine grained virtualization and API configuration.


Another important issue is reliability and fault tolerance of hypervisors. Different mechanisms and procedures need to be defined to recover faults and failures of hypervisors. A single hypervisor may not be sufficient to manage a large number of vSDN elements. To overcome scalability issue, distributed architecture of hypervisors can also be an interesting research area. Similar to controller placement, hypervisors placement plays a significant role in overall system. Research in optimizing this placement is yet another challenge
. 
\subsection{East/Westbound Interface}
Just like northbound interface of SDN, communication between different controllers is not established by a universally accepted protocol. Communication interface between multiple controllers directly effects performance. Thus, a standardized API and protocol is required so network performance can be enhanced. Due to this reason, SDN deployment is difficult in large scale networks. Also, SDN controllers play a critical role in managing and monitoring the network traffic. However, multi-controller architecture still lacks in safety mechanisms as well as suspicious traffic detection.

The different types of controllers and their architectures, also introduce heterogeneity challenge. As many of the controllers have matured over time, providing a unified eastbound interface is a challenging task. Similarly, ensuring the inter-controller communication overhead to be as minimal as possible, first requires evaluation and comparison of existing solutions, and then requires the development of an efficient communication interface. These challenges are not isolated, but also effect performance of other interfaces, such as SBI which have to install flows provided by root/master controller.

Westbound interface with traditional networks may need software implementation on routers. Such implementations may translate the flow entries to routing paths, and vice versa. 

\section{Conclusion}
This paper presented a detailed and systematic survey of different types of interfaces for Software Defined Networks. Southbound interfaces between control plane and data plane have been dominated by OpenFlow. It has become a de facto industry standard, although a number of other SBI solutions are also available which are not dependent on OpenFlow. Most of research work done in SBIs is extension or improvement of OpenFlow protocol. However, it has limited application for emerging technologies such as IoT. Northbound interfaces between application plane and controller are quite different than SBIs as the purpose is to enable users to control, configure, and program the network. We present a classification of NBIs in terms of programmability, portability, controller based, and intent based solutions. Although virtualization is highly integrated in SDN, we present it as a separate functional element, and review the different interfaces which interact with hypervisors and virtual network functions in the complete network. Inter–SDN domain interfaces and SDN to traditional network interfaces have also been analyzed in detail for their different architectures and properties. Finally, we have outlined numerous research directions and challenges for future work. 
 
\bibliographystyle{IEEEtran}
\bibliography{library}
\end{document}